\documentclass[graybox]{svmult}

\usepackage{type1cm}        
\usepackage{makeidx}         
\usepackage{graphicx}        
\usepackage{multicol}        
\usepackage[bottom]{footmisc}

\usepackage{rotating}
\usepackage{geometry} 
\usepackage{epsfig}
\usepackage{enumitem}

\usepackage{amsmath}
\usepackage{amssymb}
\def\be{\begin{equation}}
\def\ee{\end{equation}}
\def\ba{\begin{eqnarray}}
\def\ea{\end{eqnarray}}

\def\bi{\begin{itemize}}
\def\ei{\end{itemize}}

\DeclareGraphicsExtensions{.pdf,png,.jpg}

\newcommand{\app}{\emph{Astropart. Phys. }}

\newcommand{\etal}{et al.}
\newcommand{\nima}{\emph{Nucl. Instrum. Methods Phys. Res., Sect. A }}

\usepackage{newtxtext}       %
\usepackage[varvw]{newtxmath}       

\makeindex             

\begin{document}

\title*{Detecting gamma-rays with moderate resolution and large field of view: Particle detector arrays and water Cherenkov technique}
\titlerunning{VHE $\gamma$-ray air shower detectors} 
\author{Michael DuVernois and Giuseppe Di Sciascio}
\authorrunning{DuVernois \& Di Sciascio} 
\institute{Michael A. DuVernois \at Wisconsin IceCube Particle Astrophysics Center (WIPAC) \& Department of Physics, University of Wisconsin--Madison,
\\
\email{duvernois@icecube.wisc.edu}
\and Giuseppe Di Sciascio \at Istituto Nazionale di Fisica Nucleare (INFN), Sezione di Roma Tor Vergata,\\ \email{giuseppe.disciascio@roma2.infn.it}}
%
%
\maketitle


\abstract{}
The Earth is continuously bombarded by cosmic rays and gamma rays extending over an immense range of energies. Discovered in 1912 by Victor Hess, the cosmic radiation has been studied from balloons, from space, from the ground, and from underground. The resulting fields of cosmic ray astrophysics (focused on the charged particles), gamma-ray astrophysics, and neutrino astrophysics have diverged somewhat. But for the air showers in the GeV and TeV energy ranges, the ground-based detector techniques have considerable overlaps. 

Very high energy (VHE) gamma-ray astronomy is the observational study measuring the directions, flux, energy spectra, and time variability of the sources of these gamma rays. These measurements constrain the theoretical models of the sources and their interactions between the sources and detection at Earth. With the low flux of gamma rays, and the background of charged particle cosmic rays, the distinguishing characteristic of gamma-ray air shower detectors is large size and significant photon to charge particle discrimination.

Air shower telescopes for gamma-ray astronomy consist of an array of detectors capable of measuring the passage of particles through the array elements. To maximize signal at energies of a TeV or so, the array needs to be built at high altitude as the maximum number of shower particles is high in the atmosphere. These detectors have included sparse arrays of shower counters, dense arrays of scintillators or resistive plate counters (RPC), buried muon detectors in concert with surface detectors, or many-interaction-deep Water Cherenkov Detectors (WCD).

In general these detectors are sensitive over a large field of view, the whole of the sky is a typical sensitivity and perhaps 2/3 of the sky selected for clean analysis, but with only moderate resolution in energy, typically due to shower-to-shower fluctuations and the intrinsic sampling of the detector. These telescopes though, operate continuously, despite weather, moonlight, day or night, and without needing to be pointed to a specific target for essentially a 100\% duty cycle. In this chapter we will examine the performance and characteristics of such detectors. These are contrasted with the Imaging Air Cherenkov Telescopes which also operate in this energy range, and both current and future proposed experiments are described.

\section{Introduction}
\label{intro}


The large energy range that can be investigated in gamma-ray astronomy ($\approx$MeV $\to$ PeV) require different detection techniques and implies a great variety of generation phenomena.
The experimental techniques that can be used are determined by the properties of the gamma radiation and by the huge background of Cosmic Rays (CRs):
\begin{enumerate}
\item[(1)] The $\gamma$-ray flux is very small ($\le 10^{-3}$ with respect to the background of CRs detected in a $1^{\circ}$ angle around the direction of the source) and rapidly decreasing with increasing energy. All the known sources exhibit a power-law energy spectrum:
\begin{equation}
    \frac{dN}{dE} \propto E^{-\gamma}
\end{equation}
with an index $\gamma \sim$2--3. Small detectors onboard  satellites (e.g., Fermi Gamma Ray Observatory) allow observations up to around ten GeV. To detect the low flux at higher energies it is necessary to build up ground-based detectors where is possible to have larger effective collecting areas. 
\item[(2)] The Earth's atmosphere is opaque to $\gamma-$rays being about 28 radiation lengths thick at sea level. Therefore, $\gamma-$rays cannot be directly observed by ground-based detectors.
 \item[(3)] The isotropic cosmic ray flux forms a huge background exceeding by many orders of magnitude even the strongest steady photon flux. It consists largely of protons and Helium nuclei.
\end{enumerate}
It is important to consider the atmosphere also as part of the extended detector system, with atmospheric monitoring especially critical to both telescope measurements but also affecting air shower ground particle measurements.

\subsection{Ground-based detection}
The ground-based detection of VHE photons is indirect: the nature, direction, and energy of the primary particle have to be inferred from the measurable properties of the secondary particles (in the case of shower arrays) or of the Cherenkov flash in the atmosphere (for Cherenkov telescopes).
Gamma rays interact electromagnetically, producing an electron/positron pair. The mean free path of photons for pair production is almost the same as the electron radiation length, $X_0\sim$37 g/cm$^{-2}$ in air.
These secondary particles yield a new generation of $\gamma$-rays through bremsstrahlung, starting an electromagnetic Extensive Air Shower (EAS). 
The properties of electromagnetic cascades are well established for many decades \cite{heitler,matthews}, even for very small showers from low energy, few GeV, primaries \cite{epas1,epas2}.

\begin{figure}[ht!]
\begin{center}
\includegraphics[width=0.85\textwidth]{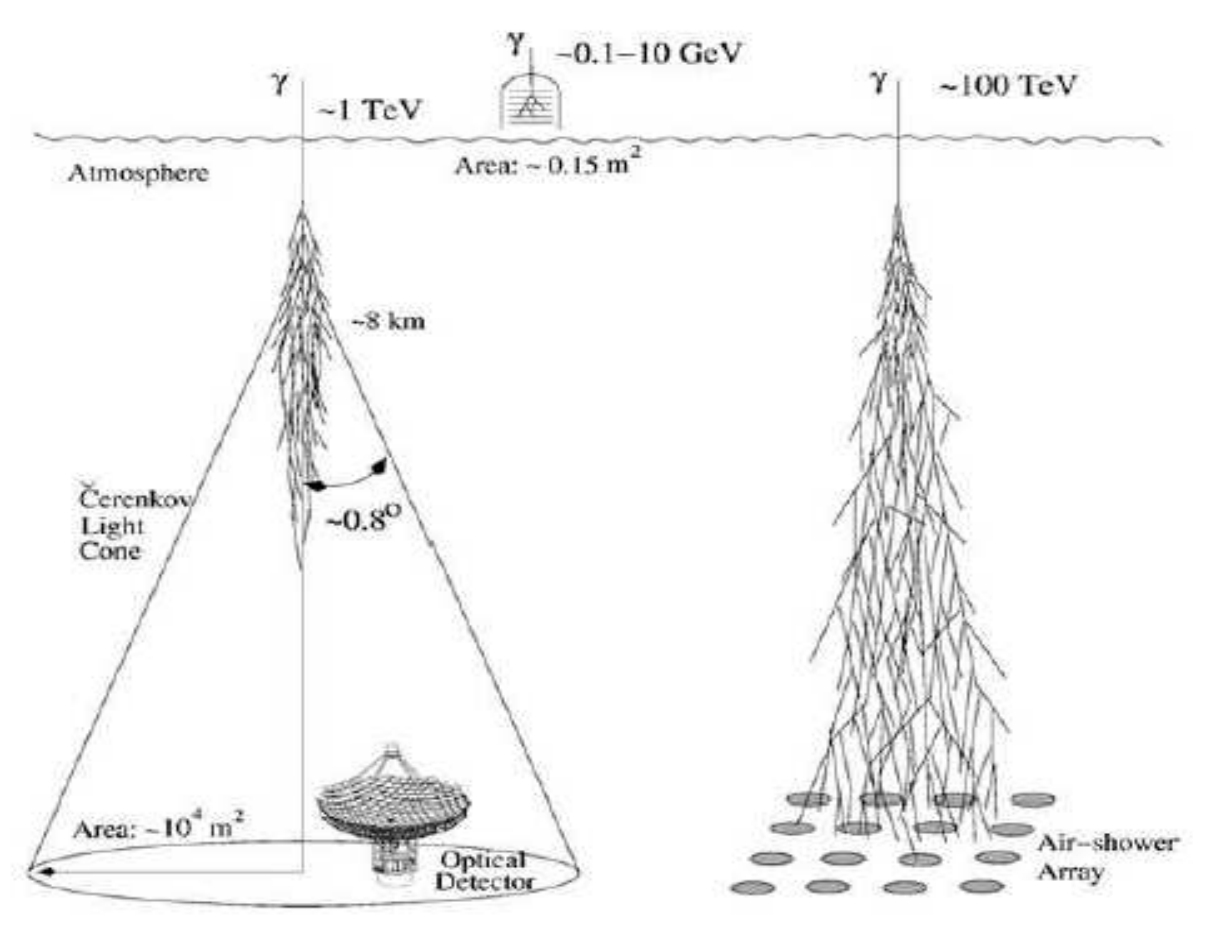}
\end{center}
\caption{Experimental techniques used for the detection of high energy $\gamma-$rays from ground. IACT on the left and air shower array on the right.}
\label{fig:chere-array}
\end{figure}

The detection of air showers from the ground is carried out by means of two different experimental techniques, both exploiting the atmosphere as a calorimeter  (Fig. \ref{fig:chere-array}):
\begin{itemize}
\item An active inhomogeneous calorimeter, with the detection of the Cherenkov light produced in air by means of telescopes.
\item A sampling calorimeter, sampling the number of secondary charged particles only at ground level (in practice, often a high altitude above sea level). 
\end{itemize}
%

Unlike Cherenkov telescopes, the detectors of shower arrays directly exploit secondary particles that reach the ground sampling their lateral distribution.
Typically they consist of a large number of charged particle detectors, usually scintillation counters, Resistive Plate Chambers (RPCs) or water Cherenkov tanks, spread over an area of 10$^4$ - 10$^5$ m$^2$ with a spacing of 10--20~m. The tail particles of the shower are sampled at a single depth (the observational level) and at fixed distances from the shower core position. In High Energy Physics language an array is a so-called \emph{Tail Catcher Sampling Calorimeter}. 

The key observables in all air shower arrays are the local shower particle densities and the secondary particle arrival times 
with which to reconstruct the shower arrival direction, the energy, and the kind of the primary particle.
The resolutions of these measurements are limited by the large shower-to-shower fluctuations mainly due to the depth of the first interaction, which fluctuates by 1 radiation length ($\sim$37 g/cm$^2$) for electromagnetic particles and by 1 interaction length ($\sim$90 g/cm$^2$) for protons.

From an experimental point of view, the sampling of secondary particles at ground can be realized  with two different approaches
\begin{enumerate}
\item[(1)] Particle Counting. A measurement is carried out with thin ($\ll$ 1 radiation length) counters providing a signal proportional to the number of charged particles (as an example, plastic scintillators or RPCs). The typical detection threshold is in the keV energy range.
\item[(2)] Calorimetry. A signal proportional to the total incident energy of electromagnetic particles is collected by a thick (many radiation lengths) detector. An example is a detector constituted by many radiation lengths of water to exploit the Cherenkov emission of secondary shower particles. The Cherenkov threshold for electrons in water is 0.8~MeV and the light yield $\approx$320 photons/cm or $\approx$160 photons/MeV emitted at 41$^\circ$.
\end{enumerate}
The critical parameters of a detector are the time and the amplitude resolutions.
The direction of the incoming primary particle is reconstructed with the \emph{time of flight method}, making use of the relative times at which the individual detection units are triggered by the shower front. 
The time resolution can affect the angular resolution of the apparatus if it is not comparable with the rising edge of the shower front ($\sim$ns). Time resolution is the convolution of the shower front leading edge, the photodetector transit-time spread, the timing jitter in the trigger system, and the finite resolution of the data acquisition. 

Sampling fluctuations are typically very large, therefore a modest amplitude resolution ($\sim$30\%) is required. A large dynamic range is on the contrary needed due to the strong dependence of the signal from the shower core distance. 
In general, not all detectors allow an optimization of both observables at the same time. As an example, in a water Cherenkov detector, reflecting inner surface allow a very good calorimetric measurement but with a narrower arrival time distribution.

Generally, the instrumented area $A$ determines the rate of high energy events recorded, i.e. the maximum energy via limited statistics. The grid distance $d$ determines the low energy threshold (small energy showers are lost in the gap between detectors) and the quality of the shower sampling. The particular kind of detector (scintillator, RPC, or water tank) determines the detail of measurement (efficiency, resolutions, energy threshold, granularity of the read-out) and impact on the cost per detector $C_d$. In principle, best physics requires large area $A$, small distance $d$ and high quality of the sampling. But the cost of an array increases with $C_d\cdot A/d^2$, therefore a compromise is always needed. 
The total sensitive area of a classical array is less than 1\% of the total enclosed area. This results in a high degree of uncertainty in the reconstruction due to sampling fluctuations which add to the shower fluctuations.
The sparse sampling sets the energy threshold and determines a poor energy resolution ($\sim$100\%). 
Also the angular resolution is limited by the shower fluctuations.

Denser arrays (see discussion of the Milagro, HAWC, and LHAASO water detectors elsewhere) can reduce the sampling fluctuations significantly while also improving the chances of catching muons in the air shower. This comes at a cost of reduced total detector area which in practice has been compensated for with a less dense outer (``outrigger'' in Milagro and HAWC parlance) detector array added onto the dense inner array. 

In addition dependence of threshold and reconstruction capabilities on the zenith angle is higher than for Cherenkov telescopes: since the active area is horizontal, its projection onto a plane perpendicular to the shower axis is smaller than the geometrical area\footnote{Cherenkov telescopes point the source therefore their area is always orthogonal to the shower axis. The degradation of the detector performance with zenith is simply due to the increasing thickness of atmosphere.}. On the other hand, EAS arrays have a large field of view ($\sim$2 sr) and a 100\% duty cycle. These characteristics give them the capability to serve as all-sky monitors, important to detect flaring, or transient, gamma emissions.
The main characteristics and differences between air shower arrays and Cherenkov telescopes are summarized in Table \ref{tab:one}.

%
\begin{table}[h]
\label{tab:one}
\vspace{0.1cm}
\centering
\caption{{\bf Air shower arrays compared to Imaging Air Cherenkov Telescopes (IACT)}}
\vspace{0.2cm}
\begin{tabular}{|c|c|c|}
\hline
\hline
  &  Air Shower Array & IACT  \\
\hline
 \hline
Energy Threshold  &  Low ($\approx$100 GeV) & Very low ($\approx$10 GeV)  \\
\hline
Max Energy  &  Very High ($\approx$10$^{15}$ eV) & Limited ($\approx$100 TeV)  \\
\hline
Field of View  &  Very Large ($\approx$2 sr) & Limited ($\approx$5 deg)  \\
\hline
Duty-cycle  &  Very large ($\approx$100\%) & Very Small ($\approx$15\%)  \\
\hline
Energy Resolution  &  Modest ($\approx$100 - 20\%) & Good ($\approx$15\%)  \\
\hline
Background rejection  &  Good ($>$80\%) & Excellent ($>$99\%)  \\
\hline
Angular Resolution  &  Good ($\approx$1-0.2 deg) & Excellent ($\approx$0.05 deg)  \\
\hline
Zenith Angle dependence  &  Very Strong ($\approx$cos$\theta^{-(6-7)}$) & Small ($\approx$os$\theta^{-2.7}$)  \\
\hline
Effective Area  &  Shrinks with zenith angle & Increases with zenith angle \\
\hline
\hline
\end{tabular} 
\end{table}

\subsection{Air shower physics}
High energy cosmic rays or gamma rays collide with the nucleus of some atmospheric gas, high in the atmosphere. The resulting collision produces additional high energy particles. These also collide with air nuclei and each additional interaction adds to the growing particle cascade. That is $N_{part} \propto 2^{depth}$ as the shower develops. These particles include neutral pions, which decay immediately to a pair of gamma rays, with the gamma rays producing $e^\pm$ pairs near other nuclei. Electrons and positrons regenerate gamma rays via bremstrahlung, building up the electromagnetic cascade. 

\subsubsection{Simplified treatment}
Decays and interactions in the atmosphere cause a $N_{part} \propto e^{-depth}$ scaling. Thus there is a point in depth of maximum number at which an air shower reaches a maximum size $N_{max}$ which is approximately E/(1.6~GeV) depending on some details such as incoming species and the hadronic interactions. Gamma-ray initiated showers are purely electromagnetic, so let's consider that subset of cosmic-ray air showers.

In the electromagnetic shower $\pi^0$ mesons decay into gamma rays, with gammas converting in to $e^\pm$ pairs, which produce new gammas by bremsstrahlung. The radiation length $X_0$ is the grammage path length ($\int$density(x) dx) in which their energies attenuate by a factor of 1/e. In air, the radiation length $X_0$ is about 37~g/cm$^2$. One can picture the cascade as a series of generations. For each generation every existing gamma ray converts to electron-positron pair, while each existing electron or positron produces a new gamma ray in addition to itself. 

Each generation doubles the number of cascade particles. This continues until the average particle energy is reduced to a critical energy scale at which the charged particles lose energy to ionization in less than a radiation length. Ionizing energy loss is about 2.2~MeV/g/cm$^2$, so the critical energy is (2.2~MeV/g/cm$^2$) $\times$ (37~g/cm$^2$) = 81~MeV. At this point, the shower size is $N_{max}$ = E/E$_c$. The number of generations n needed to reach this maximum size depends on the total energy E. At maximum, 2$^n$ = N$_{max}$ = E/E$_c$, so n = ln(E/E$_c$)/ln(2). The maximum size occurs at a (slant) depth of X$_{max}$ = n $\times$ X$_0$ $\times$ ln(2) = X$_0$ $\times$ ln(E/E$_c$) along the shower axis.

The number of particles as a function of depth is typically called the electromagnetic longitudinal profile, and is described by the Greisen formula:

\begin{equation}
N_e = \frac{0.31}{\sqrt{T_{max}}} e^T s^{-3T/s}.
\end{equation}

Here T is the atmosphere depth measured in radiation length X/X$_0$ in slant distance, T$_{max}$=ln(E/E$_c$), and $s$ is the shower age s=3T/(T+2T$_{max}$). Shower maximum is at shower age s=1 \cite{greisen}.

Hadronic interactions, from a charged cosmic ray, additionally produce charged pions, $\pi^\pm$ and muons, $\mu^\pm$, in addition to forward nuclear fragments. For a good reference on the structure of hadronic air showers, see here \cite{hadronic}. The secondary muons at the ground therefore form an observable consequence of a charged primary cosmic ray, and can be used to as a veto to suppress the (dominant) charged cosmic ray flux.

\subsubsection{Adding complexity to the air shower model}
The simplified form of the air shower physics described above gives us some qualitative understanding, but in practice, the air shower development is studied via Monte Carlo-based simulations using either CORSIKA or AIRES to handle the overall atmospheric simulation and some combination of GEANT and hadronic simulation codes such as QGSJET for the underlying micro-physics \cite{corsika,aires,geant,qgsjet}. This allows for the shower-to-shower fluctuations to be taken into account and also addresses the atmosphere as a detector element (see below), timing and positions of ground-level particles, and the shape of the shower front which is critical for timing.

\begin{figure}[bh]
    \centering
    \vspace{-1.7cm}
    \includegraphics[width=9cm]{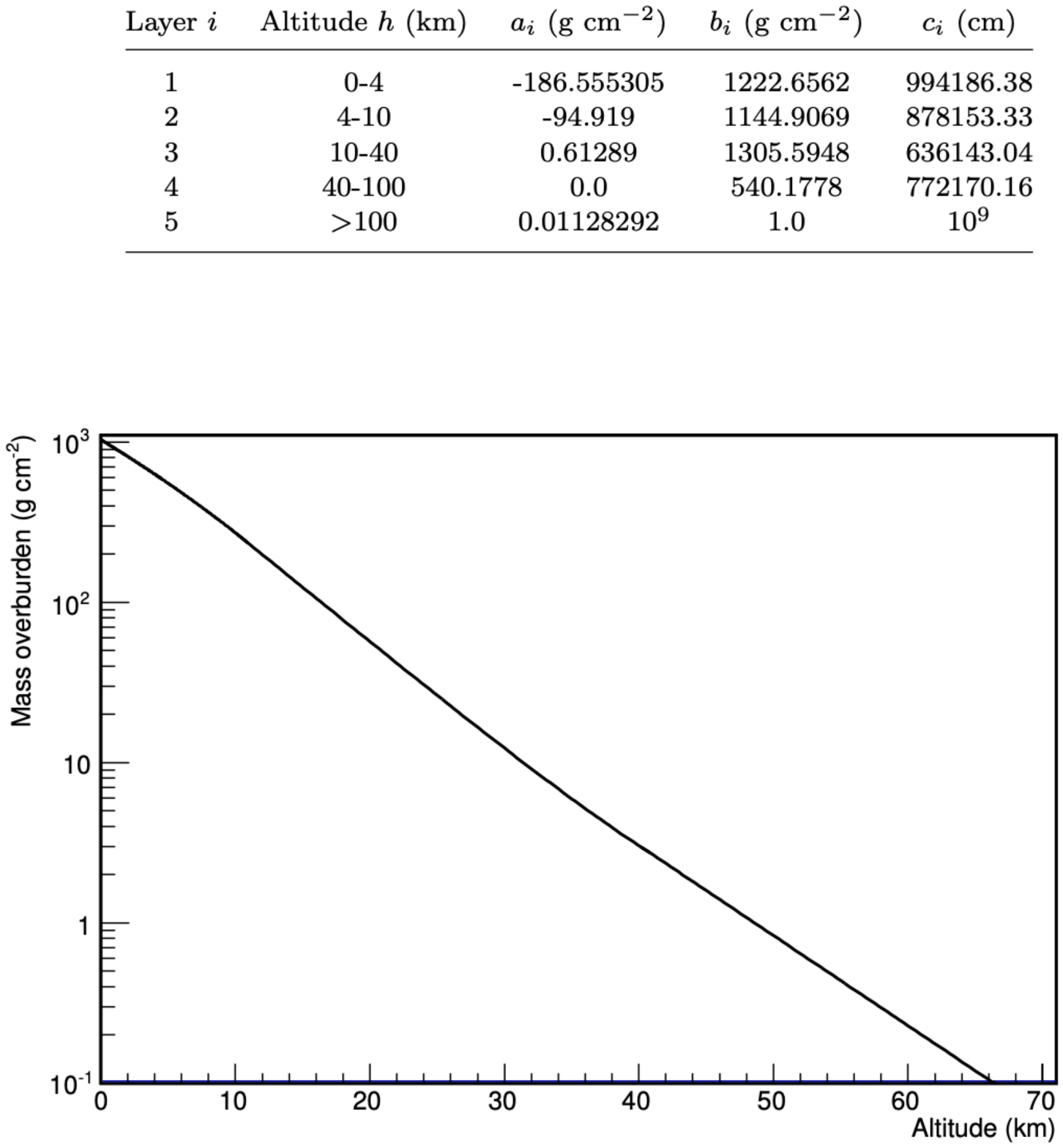}
    \vspace{-2cm}
    \caption{U. S. Standard Atmosphere mass overburden as function of altitude and parameterization of the atmospheric overburden in five layers.}
    \label{fig:atmos}
\end{figure}

The atmosphere is composed of about 78\% N$_2$, 21\% O$_2$, and 1\% Ar, with an average atomic mass of $\sim$14.6. This average atomic mass is generally used as the target ``nucleus.'' The starting point for an atmospheric model is generally the U.S. Standard Atmosphere developed by NOAA, NASA, and the US Air Force \cite{standard}. This has a typical parameterization into five layers with simple coefficients \cite{corsika,wcd} and shown in Fig. \ref{fig:atmos}. This parameterization functions for the first 4 layers are exponential:
\begin{equation}
    X(h) = a_i + b_i \times e^{-h/c_i}
\end{equation}
and the 5th layer is linear in altitude and overburden:
\begin{equation}
    X(h) = a_5 - b_5 \times \frac{h}{c_5}.
\end{equation}
The U. S. Standard Atmosphere is approximately valid for most low altitude sites around the world with minor deviations, at higher altitudes, and climatically complicated sites such as the South Pole, there are modified forms appropriate to the local sites.

\subsection{Example experiments}
The earliest air shower arrays used for gamma-ray astronomy were composed of small plastic scintillator panels distributed over a large area. With low active area fractions of less than 1\% of the enclosed area, these arrays had high energy thresholds near 100~TeV. The CYGNUS \cite{cygnus} and CASA \cite{casa} arrays were the largest of this sort of instrument. No evidence of sources were seen in these experiments.

We will look more closely at two experiments of the next generation of gamma-ray observatories using the experimental approaches mentioned above, but with a goal of lowering the detector thresholds. These two different ground-based TeV survey instruments and their techniques:
\begin{enumerate}
\item[(1)] Water Cherenkov (Milagro) \cite{milagro}.
\item[(2)] Resistive Plate Chambers (ARGO-YBJ) \cite{disciascio-rev}. 
\end{enumerate}

The Milagro detector consisted of a large central water reservoir (60$\times$80 m$^2$), which operated between 2000 and 2008 in New Mexico (36$^{\circ}$ N, 107$^{\circ}$ W), at an altitude of 2630~m above sea level (a.s.l.) \cite{milagro}. The reservoir was covered with a light-tight barrier, and instrumented with 2 layers of 8'' PMTs to improve the detection of muons. In 2004, an array of 175 small tanks was added, irregularly spread over an area of 200$\times$200 m$^2$ around the central reservoir. With this array, the Milagro collaboration developed analysis techniques for CR background discrimination (gamma-hadron separation).

The Tibet AS$\gamma$ experiment utilized a sparse array of plastic scintillators with an energy threshold higher than $\approx$10 TeV \cite{amenomori2015}.

\begin{figure}[!t]
\centerline{\includegraphics[width=0.9\textwidth]{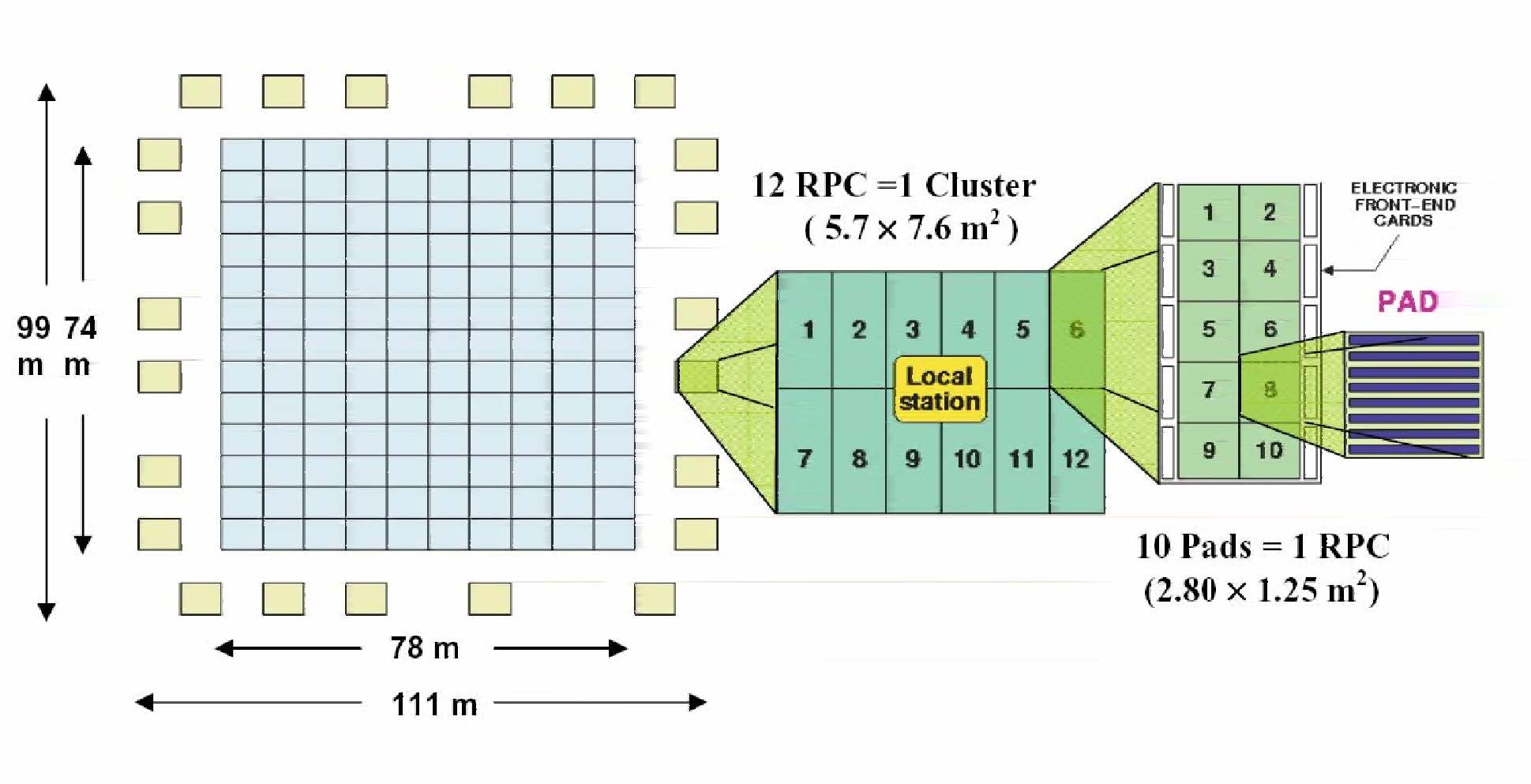}}
  \caption{Layout of the ARGO-YBJ experiment (see text for a description of the detector).}
  \label{fig:fig-01}
\end{figure}

The ARGO-YBJ experiment, located at the Yangbajing Cosmic Ray Observatory (Tibet, PR China, 4300~m a.s.l., 606 g/cm$^2$), is constituted by a central carpet $\sim$74$\times$78 m$^2$, made of a single layer of Resistive Plate Chambers (RPCs) with $\sim$93$\%$ of active area, enclosed by a guard ring partially instrumented ($\sim$20$\%$) up to $\sim$100$\times$110 m$^2$. The apparatus has a modular structure, the basic data acquisition element being a cluster (5.7$\times$7.6 m$^2$), made of 12 RPCs (2.85$\times$1.23 m$^2$ each). Each chamber is read by 80 external strips of 6.75$\times$61.80 cm$^2$ (the spatial pixels), logically organized in 10 independent pads of 55.6$\times$61.8 cm$^2$ which represent the time pixels of the detector \cite{aielli06}. 
The readout of 18,360 pads and 146,880 strips is the experimental output of the detector. 
In addition, in order to extend the dynamical range up to PeV energies, each chamber is equipped with two large size pads (139$\times$123 cm$^2$) to collect the total charge developed by the particles hitting the detector \cite{argo-bigpad}.
The RPCs are operated in streamer mode by using a gas mixture (Ar 15\%, Isobutane 10\%, TetraFluoroEthane 75\%) optimized for high altitude operation \cite{bacci00}. The high voltage settled at 7.2 kV ensures an overall efficiency of about 96\% \cite{aielli09a}.
The central carpet contains 130 clusters and the full detector is composed of 153 clusters for a total active surface of $\sim$6700 m$^2$ (Fig. \ref{fig:fig-01}). The total instrumented area is $\sim$11,000 m$^2$.
%
%

Because of the small pixel size, the detector is able to record events with a particle density exceeding 0.003 particles $\cdot$ m$^{-2}$, keeping good linearity up to a core density of about 15 particles $\cdot$ m$^{-2}$.
The median energy of the first multiplicity bin (20-40 fired pads) for photons with a Crab-like energy spectrum is $\sim$340 GeV \cite{argo-crab}.
The granularity of the read-out at centimeter level and a noise of accidental coincidences of 380 Hz/pad allowed to sample events with only 20 fired pads, out of 15,000, with a noise-free topological-based trigger logic.

The benefits in the use of RPCs in ARGO-YBJ are \cite{disciascio-rev,bartoli2011}: 
\begin{enumerate}
    \item[(1)] high efficiency detection of low energy showers by means of the high density sampling of the central carpet (the detection efficiency of 100 GeV photon-induced events is $\approx$50\% in the first multiplicity bin); 
    \item[(2)] unprecedented wide energy range investigated by means of the digital/charge read-outs ($\sim$300 GeV $\to$ 10 PeV);
    \item[(3)] good angular resolution ($\sigma_{\theta}\approx 1.66^{\circ}$ at the threshold, without any lead layer on top of the RPCs) and unprecedented details in the core region by means of the high granularity of the different read-outs.
\end{enumerate}

RPCs allowed one to also study charged cosmic-ray physics (energy spectrum, elemental composition, and anisotropy) up to about 10 PeV.
By contrast, the capability of water Cherenkov facilities in extending the energy range to PeV and in selecting different primary masses must be still investigated.

In both experiments (Milagro and ARGO-YBJ) the limited capability to discriminate the background was mainly due to the small dimensions of the central detectors (pond and carpet). In fact, in the new experiments HAWC \cite{hawc1} and LHAASO \cite{lhaaso1,cao2021nat}, the discrimination of the CR background is made studying shower characteristics far from the shower core (at distances R$>$ 40 m from the core position, the dimension of the Milagro pond and ARGO-YBJ carpet). 

\subsection{HAWC}

%
\begin{figure}[bh]
\vspace{-0.2cm}
\centerline{\includegraphics[width=\textwidth]{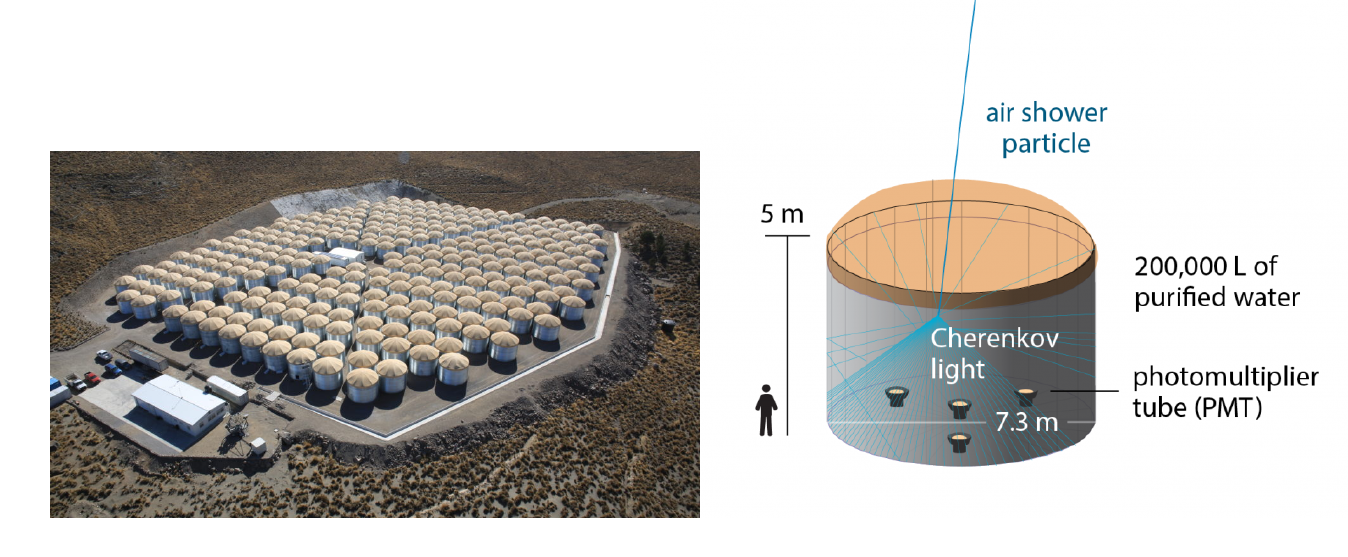} }
\caption{Layout of the HAWC experiment as seen from aerial photograph. The sketch of a water tank is shown on the right side.} 
\label{fig:hawc-layout}       
\end{figure}
%
The Milagro results, as well as the potential for continuous monitoring of a large fraction of the sky, have motivated the construction of larger EAS detectors like the High-Altitude Water Cherenkov Observatory (HAWC). 
HAWC is located on the Sierra Negra volcano in central Mexico at an elevation of 4100 m a.s.l.  (see Fig. \ref{fig:hawc-layout}).
It consists of an array of 300 water Cherenkov detectors made from 5 m high, 7.32 m diameter, water storage tanks covering an instrumented area of about 22,000 m$^2$. Water in the tanks totals about 55 kilotons. 
Four upward-facing photomultiplier tubes (PMTs) are mounted at the bottom of each tank: a single 10'' High Quantum Efficiency (HQE) PMT positioned at the center, and three 8'' PMTs formerly from Milagro positioned halfway between the tank center and rim. 
The central PMT has roughly twice the sensitivity of the outer PMTs, due to its superior quantum efficiency and larger size. The WCDs are filled to a depth of 4.5 m, with 4.0 m (more than 10 radiation lengths) of water above the PMTs. This large depth guarantees that the electromagnetic particles in the air shower are fully absorbed by the HAWC detector, well above the PMT level, so that the detector itself acts as an electromagnetic calorimeter providing an accurate measurement of electromagnetic energy deposition \cite{hawc1,hawc12,hawc2}. HAWC operates with a two stage triggering system: PMT pulses above threshold generate signal edges for two levels of small and large events; and all of these digital edges are brought together in software to form interesting event triggers. The purely software second level trigger allows for significant flexibility in addressing horizontal showers, and potentially lowered thresholds in certain directions in addition to facilitating the basic air shower trigger \cite{hawcnim}.

\subsection{LHAASO}
A new project, developed starting from the experience of ARGO-YBJ and the Milagro and HAWC water tanks, is LHAASO \cite{cao2021nat}. 
The experiment is strategically built to study with unprecedented sensitivity the energy spectrum, the elemental composition and the anisotropy of CRs in the energy range between 10$^{12}$ and 10$^{17}$ eV, as well as to act simultaneously as a wide aperture ($\sim$2 sr), continuossly-operated gamma-ray telescope in the energy range between 10$^{11}$ and $10^{15}$ eV.

\begin{figure}[ht!]
\centerline{\includegraphics[width=0.8\textwidth]{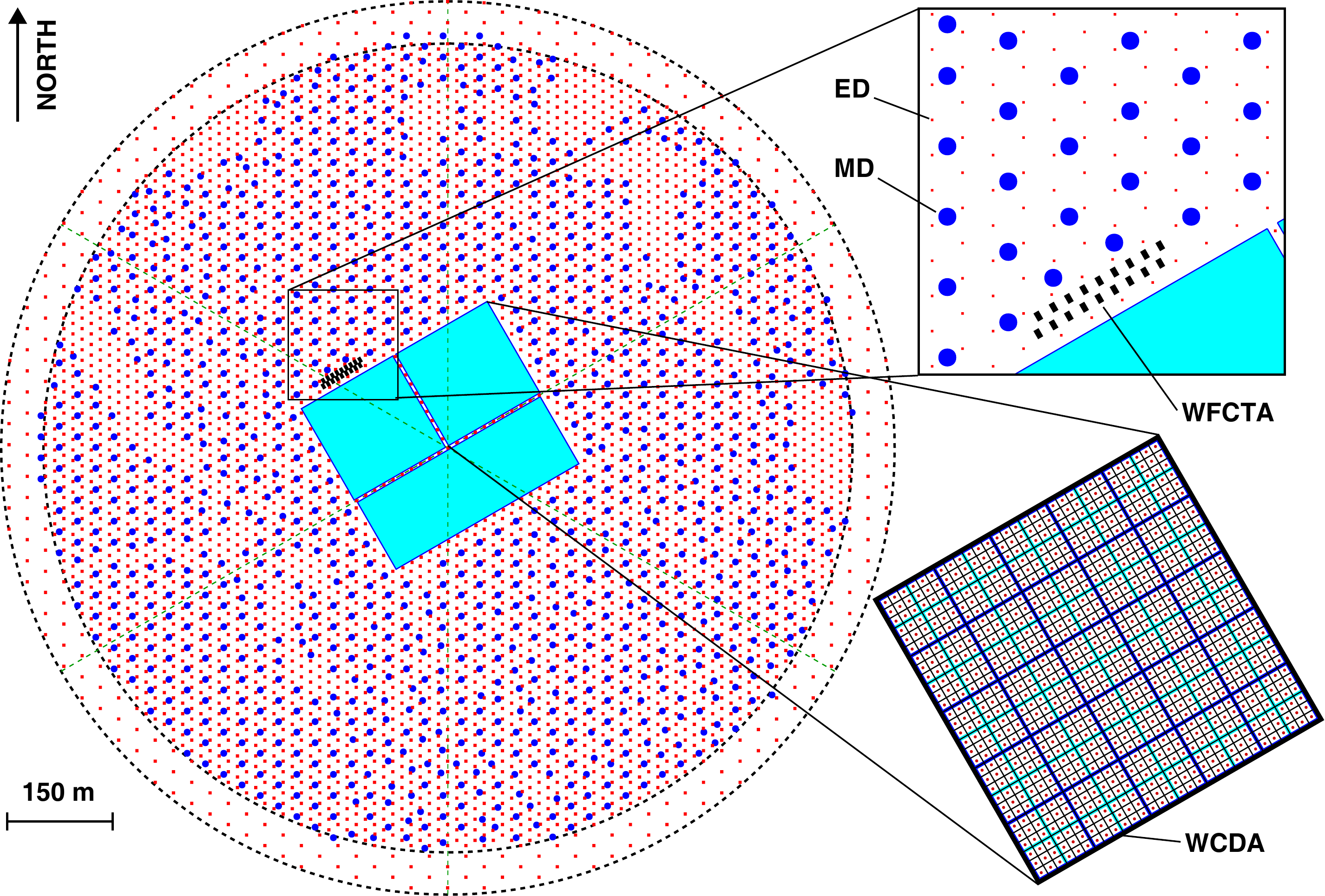} }
\caption{Layout of the LHAASO experiment. Central pond and outer particle detector array.} 
\label{fig:lhaaso-layout}       
\end{figure}

The first phase of LHAASO consists of the following major components (see Fig. \ref{fig:lhaaso-layout}):
\begin{itemize}
\item 1.3 km$^2$ array (LHAASO-KM2A) for electromagnetic particle detectors (ED) divided into two parts: a central part including 4931 scintillator detectors 1 m$^2$ each in size (15~m spacing) to cover a circular area with a radius of 575~m and an outer guard-ring instrumented with 311 EDs (30~m spacing) up to a radius of 635~m.
\item An overlapping 1~km$^2$ array of 1146 underground water Cherenkov tanks 36~m$^2$ each in size, with 30~m spacing, for muon detection (MD, total sensitive area $\sim$42,000~m$^2$).
\item A close-packed, surface water Cherenkov detector facility with a total area of about 78,000~m$^2$ (LHAASO-WCDA).
\item 18 wide field-of-view air Cherenkov telescopes (LHAASO-WFCTA).
\end{itemize}
LHAASO is located at high altitude (4410 m a.s.l., 600 g/cm$^2$, 29$^{\circ}$ 21' 31'' N, 100$^{\circ}$ 08'15'' E) in the Daochen site, Sichuan province, P.R. China. 
%
%

In Table \ref{tab:array-summary} and Table \ref{tab:array-summary-muon} the main characteristics of a selection of air shower arrays operated in the last two decades to study gamma-ray astronomy and galactic cosmic-ray physics from ground are summarized. 
The atmospheric depths of the arrays, the main detectors used, the energy range investigated, the sensitive areas of electromagnetic and muon detectors, the instrumented areas and the coverage (i.e., the ratio between sensitive and instrumented areas) are reported.

The depth in the atmosphere (altitude of the site) is crucial to fix the energy threshold, the energy resolution and the impact of shower-to-shower fluctuations. 
As can be seen, the new experiment LHAASO will operate with a coverage of $\sim$0.5\% over a 1 km$^2$ area.
The sensitive area of the muon detectors is unprecedented and about 17 times larger than CASA-MIA, with a coverage of about 5\% over 1 km$^2$.
%
\begin{table}[ht]
\caption{\label{tab:array-summary} Characteristics of a selection of air shower arrays}
\footnotesize
\begin{center}
\begin{tabular}{lllllll}
\hline
\hline
  Experiment &  g/cm$^2$ & Detector & $\Delta$E & e.m.\ sens. & Instr. & Coverage \\
                     &                  &                &       (eV)       &         area (m$^2$)             & area (m$^2$) & \\
\hline
ARGO-YBJ \cite{bartoli2011}  & 606            & RPC/hybrid  with  & $3\times 10^{11} - 10^{16}$ & 6700 & 11,000 & 0.93 \\
                    &                   &   wide-FoV \v C Tel.   &                                            &          &  & (carpet)\\
HAWC \cite{hawc1} & 620 & Water \v C & $10^{12} - 10^{14}$ & 1.2$\times$10$^4$ & 2$\times$10$^4$ & 0.61 \\
 TIBET AS$\gamma$ \cite{amenomori2011} & 606 &  scint./burst det.  & $5\times 10^{13} - 10^{17}$ & 380 & 3.7$\times$10$^4$ & 10$^{-2}$ \\
 CASA-MIA \cite{casamia} & 860 & scint./muon & 10$^{14} - 3.5\times 10^{16}$ & 1.6$\times$10$^3$ & 2.3$\times$10$^5$ & 7$\times$10$^{-3}$ \\
KASCADE \cite{kascade} & 1020 & scint./mu/had & $2\times 10^{15} - 10^{17}$ & 5$\times$10$^2$ & 4$\times$10$^4$ & 1.2$\times$10$^{-2}$ \\
KASCADE-& 1020 & scint./mu/had  & $10^{16} - 10^{18}$ & 370 & 5$\times$10$^5$ & 7$\times$10$^{-4}$ \\
Grande \cite{kascade-grande} &  &  &  &  &  &  \\
Tunka \cite{prosin2014} & 900 & open \v C det. & 3$\times 10^{15} - 3\times 10^{18}$ & --- & 10$^6$ & --- \\
IceTop \cite{aartsen2019} & 680 & ice \v C det. & $10^{16} - 10^{18}$ & 4.2$\times$10$^2$ & 10$^6$ & 4$\times$10$^{-4}$ \\
 LHAASO \cite{cao2021nat} & 600 & Water \v C& $10^{12} - 10^{17}$ & 5.2$\times$10$^3$ & 1.3$\times$10$^6$ & 4$\times$10$^{-3}$ \\
                    &        & scint./mu/had     &                                            &          &  & \\
                    &        & wide-FoV \v C Tel.     &                                            &          &  & \\
\hline
\hline
\end{tabular}
\end{center}
\end{table}

\begin{table}[ht]
\caption{\label{tab:array-summary-muon} Characteristics of muon detectors operated in a selection of shower arrays}
\begin{center}
\begin{tabular}{ccccc}
\hline
\hline
 Experiment &Altitude & $\mu$ Sensitive Area & Instrumented Area & Coverage \\
  & (m) & (m$^2$) & (m$^2$) & \\
\hline
LHAASO & 4410 & 4.2$\times$10$^4$ & 10$^6$ & 4.4$\times$10$^{-2}$ \\
 TIBET AS$\gamma$ & 4300 & 4.5$\times$10$^3$ & 3.7$\times$10$^4$ & 1.2$\times$10$^{-1}$ \\
 KASCADE & 110 & 6$\times$10$^2$ & 4$\times$10$^4$ & 1.5$\times$10$^{-2}$ \\
 CASA-MIA & 1450 & 2.5$\times$10$^3$ & 2.3$\times$10$^5$ & 1.1$\times$10$^{-2}$ \\
\hline
\hline
\end{tabular} 
\end{center}
\end{table}
%

\section{Detector performance}
To understand the overall performance, and tradeoffs in design, we look here at some measures of a successful TeV-scale gamma-ray observatory:
\begin{enumerate}
\item[(1)] Sensitivity to $\gamma$-point source
\item[(2)] Energy threshold
\item[(3)] Trigger relative efficiency
\item[(4)] Angular resolution
\item[(5)] Background discrimination
\end{enumerate}

The recent construction of the first portions of LHAASO have shown the direction taken by that collaboration in light of the previous generations of experiments. LHAASO is using multiple technologies, greatly expanding the scale of the detector, and taking advantage of the multiple technologies to offer different ``handles'' on managing the cosmic-ray background. We'll also look at prospects for the future, which include the SWGO (Southern Wide-field Gamma-ray Observatory) effort, currently in the design process, and actively managing these different detector performance metrics for potential high altitude sites in South America to give a southern hemisphere sensitivity at TeV gamma-ray energies. For one approach to this trade study, see this recent work \cite{harm}.

Outside of the air shower measurements, an important feature of the wide field of view gamma-ray experiments is their sensitivity to transient, and multi-messenger, events. The uptime of the HAWC detector has been better than 98\% with an instantaneous coverage of about 1/3 of the sky. As astronomy becomes increasingly focused on transient events, always-on, no-pointing-required, observations are especially important at high energies which for many phenomena are peaked early in the emission. Additionally, following up from neutrino or gravitational-wave observations can be done using the archived data of these detectors rather than an attempt at quickly re-positioning and targeting the follow-on signals. Multi-messenger networks of communications, such as AMON \cite{amon}, allow for the distribution of observations of different messengers, from very different styles of telescopes, in near real-time.

\subsection{Sensitivity to a $\gamma$-ray point source}
 
The main drawback of ground-based measurement consists is the huge background of charged CRs.
This means that ground-based instruments detect a source as an excess of events from a certain direction over an overwhelming uniform background. Suppression of the cosmic-ray signal can be achieved by use of muon detector veto or other gamma-hadron distinguishing variables in the air shower signal. Due to the background though, the detector sensitivities are often expressed in units of standard deviations of the cosmic ray background (see for example \cite{disciascio19})
 \begin{equation}
\label{eq:mdf}
S = \frac{N_{\gamma}}{\sqrt{N_\text{bkg}}} \propto \frac{\Phi_{\gamma}}{\sqrt{\Phi_\text{bkg}}}\cdot R\cdot \sqrt{A_\text{eff}^{\gamma}}\cdot \frac{1}{\sigma_{\theta}}\cdot \sqrt{T}\cdot {Q}
\end{equation}
where $\Phi_{\gamma}$ and $\Phi_\text{bkg}$ are the integral fluxes of photon and background, $\sigma_{\theta}$ is the angular resolution, 
\begin{equation}
    R=\sqrt{A_\text{eff}^{\gamma}/A_\text{eff}^\text{bkg}},
\end{equation}
the $\gamma$/hadron relative trigger efficiency and $T$ the observation time.
The so-called \emph{Q-factor}
\begin{equation}
    Q=\frac{\epsilon_{\gamma}}{\sqrt{1-\epsilon_\text{bkg}}}
\end{equation}
represents the gain in sensitivity due to the background discrimination capability, where $\epsilon_{\gamma}$ and $\epsilon_\text{bkg}$ are the efficiencies in identifying $\gamma$-induced and background-induced showers, respectively.

For a point source the angular term to evaluate the isotropic background is given by the opening angle of the detector, i.e. the point spread function PSF ($\Delta\Omega = \Delta\Omega_{PSF} \sim \pi\, \theta^2_{PSF}$).
If we have an \emph{extended source} with a photon flux equal to that of the point source we must integrate over the extension of the source to have the same number of photons: $\Delta\Omega_{PSF}\to \Delta\Omega_{ext}$, and the background will increase.
Therefore, 
\be
S_{ext}\propto \bigg[ \frac{\Phi_{\gamma}}{\sqrt{\Phi_{bkg}}}\cdot R\cdot \sqrt{A_{eff}^{\gamma}}\cdot Q \cdot \frac{1}{\theta_{ext}} \cdot \frac{\theta_{ext}}{\theta_{PSF}}\bigg] \cdot \frac{\theta_{PSF}}{\theta_{ext}}
\ee
and
\be
S_{ext}\propto S_{point}\cdot \frac{\theta_{PSF}}{\theta_{ext}}
\ee
where $\theta_{ext}$ is the dimension of the extended source.
As it can be seen, detectors with a poor angular resolution, like shower arrays, are favoured in the extended source studies. 

Because for the integral fluxes we can write 
\begin{equation}
    \Phi_{\gamma}\sim E_{thr}^{-\gamma} \; \textrm{and} \;  \Phi_{bkg}\sim E_{thr}^{-\gamma_{bkg}}
\end{equation}
we obtain
\begin{equation}
\frac{\Phi_{\gamma}}{\sqrt{\Phi_{bkg}}}\sim E_{thr}^{-(\gamma - \gamma_{bkg}/2)} \sim E_{thr}^{-2/3}
\end{equation}
being $\gamma\sim$1.5 and $\gamma_{bkg} \sim$1.7.

Energy threshold, relative trigger probability R, angular resolution and Q-factor are the main parameters, the drives, which determine the sensitivity of a ground-based $\gamma$-ray telescope.

The wide angular acceptance of air shower arrays also leads to a somewhat complicated extended source sensitivity. In particular, extended source sensitivities of large field of view instruments are often superior compared to narrow field instruments which require multiple exposures to image the full object due to background subtraction. Difference between the HAWC and HESS, for example, Galactic plane extended sources have been noted in this area \cite{hesshawc}.

\subsection{The energy threshold}
 
The energy threshold of EAS-arrays is not well defined due to the large fluctuations affecting the number of particles at ground from shower to shower for identical primaries. The main source of fluctuations is the depth of first interaction. As a consequence, an array can be triggered by very low energy showers if the primary particles interacted by chance deeper into atmosphere than expected. On the contrary, may fail to detect high energy events when the initial interaction is unusually high in the atmosphere. Therefore, the trigger probability increases slowly with energy and is not a step function at the threshold energy $E_{thr}$.
The shower fluctuations can be reduced placing the detector at high altitude, close to the maximum of the shower development.

As a rule of thumb, we can estimate the energy threshold of an array from the effective area and trigger rate \cite{cronin95}. If we assume that the efficiency as a function of cosmic ray energy has a turn-on at an energy E$_0$, the rate is given by
\begin{equation}
R = \Phi_{CR}(\geq E_0) \times (\sim1 \text{sr})\times A 
\end{equation}
where $\Phi_{CR}$ is the integral CR flux and $A$ is the effective area of the array. The rate of cosmic rays peaks at the zenith and more than 90\% of the recorded showers lie within a cone of 1 sr around the zenith. Therefore, we assume the effective solid angle of an array equal to 1 sr. This is also the angular range over which the array has a reasonable acceptance for a source passing overhead.
According to a well-known parametrization \cite{horandel} the integral flux is given by
\begin{equation}
\Phi_{CR}(\geq E_0) = 1.3\cdot \left(\frac{E_0}{1\>\text{GeV}}\right)^{-1.66} \textrm{cm$^{-2}$s$^{-1}$sr$^{-1}$}.
\end{equation}
Solving for the energy we find 
\begin{equation}
E_0 =  1.2\cdot \left(\frac{R_{Hz}}{A_{cm^2}}\right)^{-0.6}  \textrm{GeV}.
\end{equation}
For ARGO-YBJ, a rate of 3.5 kHz and an instrumented area of 10$^8$ cm$^2$ implies E$_0\sim$ 470 GeV, in fair agreement with the median energy of the first multiplicity bin of 340 GeV.

The energy threshold of a shower array is mainly determined by the combination of altitude and coverage, i.e. the ratio between sensitive and instrumented areas.
In addition, the threshold of the particular detector and the trigger logic of the apparatus can affect the final energy threshold. An important limiting factor is the rate of accidental coincidences of the detector unit. In the ARGO-YBJ experiment the single particle counting rate is about 1 kHz/m$^2$, to be compared with the rate of single muons of about 200 Hz/m$^2$. 
In the LHAASO experiment the counting rate of a single 8'' PMT used in WCDA is about 40~kHz. HAWC has a similar hardware trigger rate of a single 8'' PMT of around 20~kHz at 1/4 SPE (single photo electron). These higher PMT rates require a larger number of channels triggered for a reconstructable event to rise above the noise floor.

%
\begin{figure}[h!]
\begin{minipage}[t]{.47\linewidth}
  \centerline{\includegraphics[width=\textwidth]{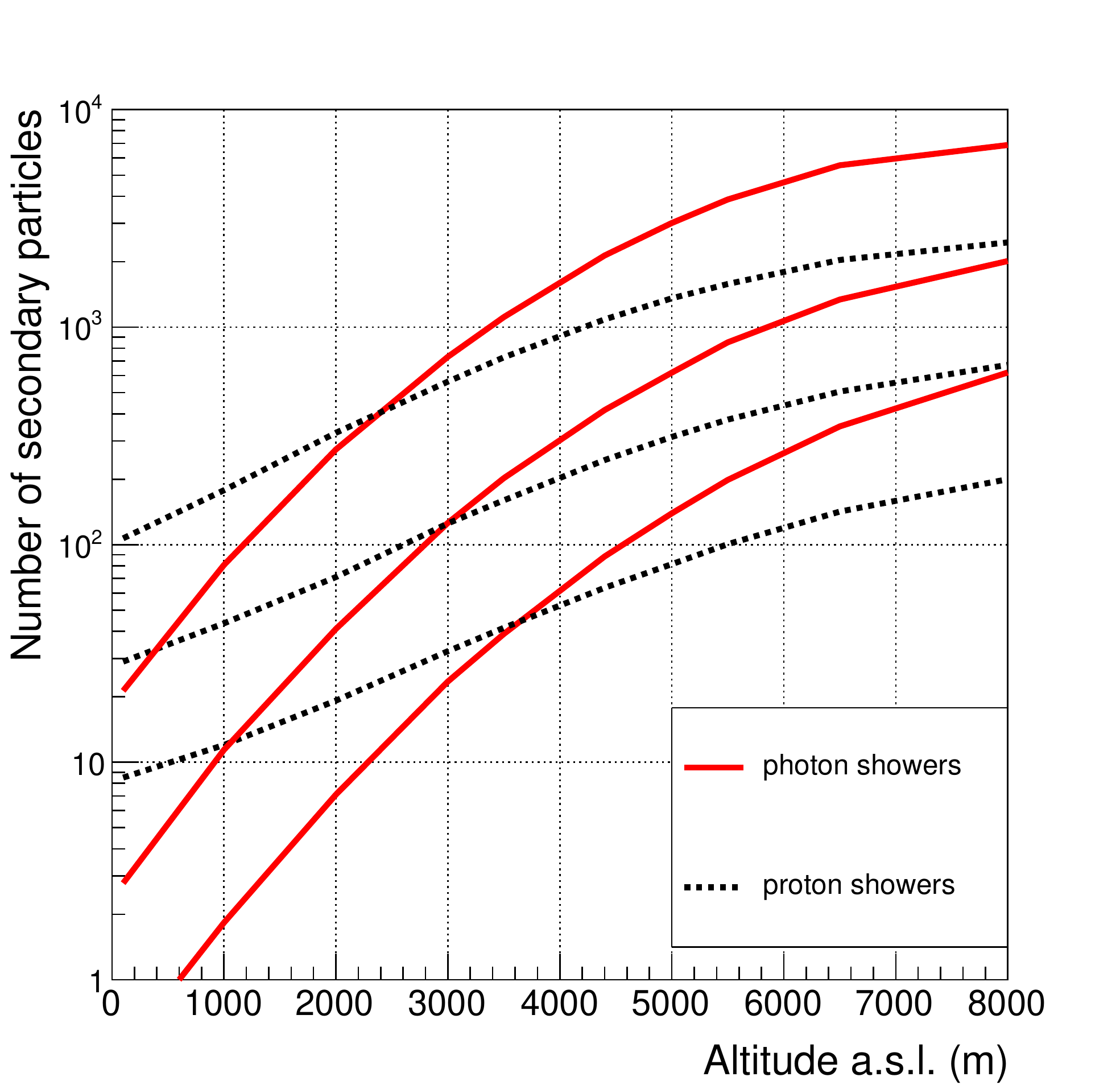} }
\end{minipage}\hfill
\begin{minipage}[t]{.47\linewidth}
  \centerline{\includegraphics[width=\textwidth]{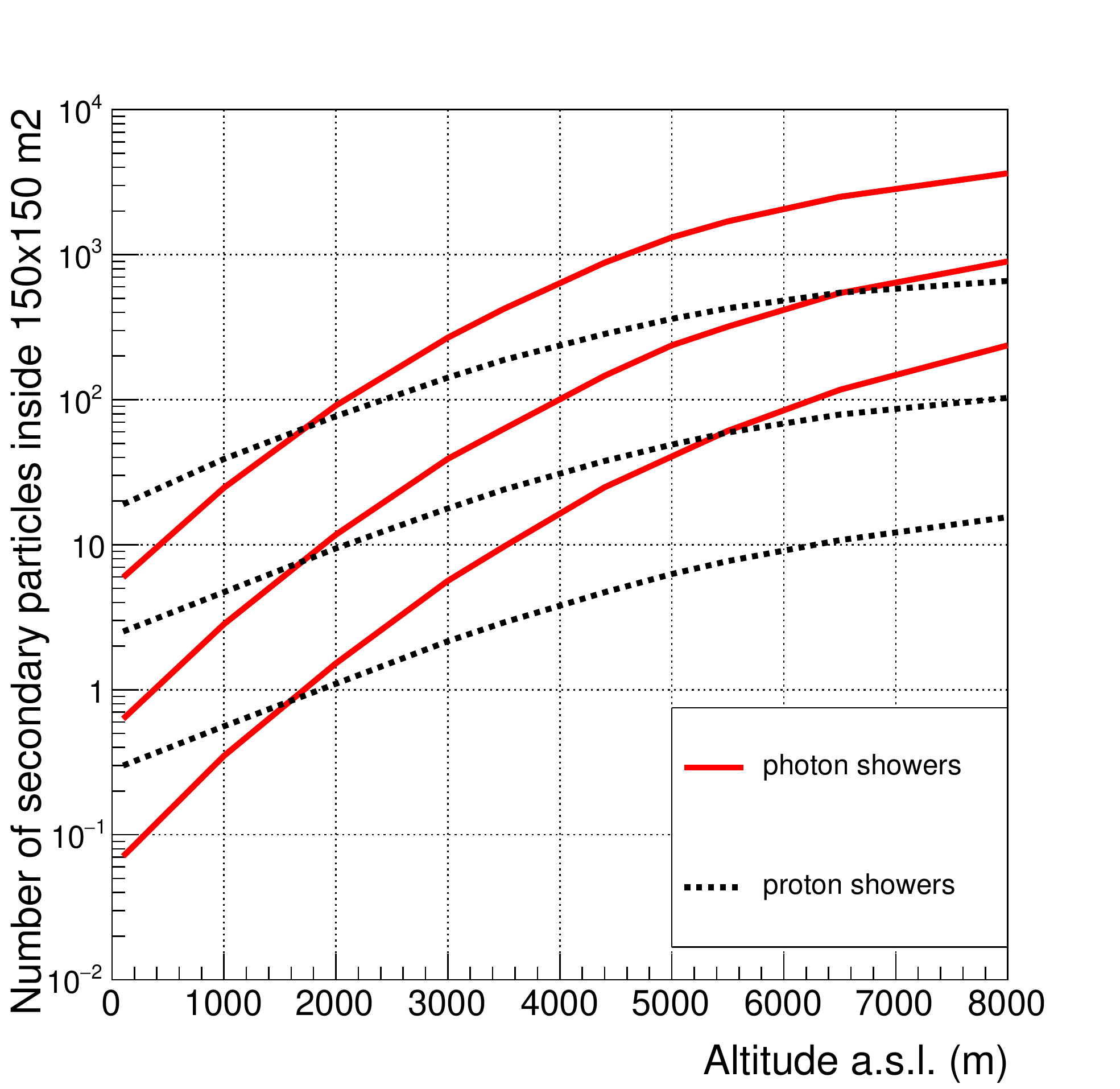} }
\end{minipage}\hfill
\caption[h]{Average number of particles (charged + photons) produced by showers induced by primary photons and protons of different energies at different observation levels. The left plot shows the total size, the right one refers to particles contained inside an area 150$\times$150 m$^2$ centered on the shower core. The plotted energies are 100, 300 \& 1000 GeV starting from the bottom \cite{disciascio-icrc2017}.} 
\label{fig:size}
\end{figure}
%
In the Fig. \ref{fig:size} the average sizes produced by showers induced by primary photons and protons of different energies at different observation levels are plotted. The left plot shows the total number of secondary particles (charged plus photons), the right one shows the number of particles contained inside an area 150$\times$150 m$^2$ centered on the shower core.
As can be seen, the number of particles in proton-induced events exceeds the number of particles in $\gamma$-induced ones at low altitudes. This implies that, in gamma-ray astronomy, the trigger probability is higher for the background than for the signal.

\begin{figure}[h!]
\begin{minipage}[t]{.47\linewidth}
  \centerline{\includegraphics[width=\textwidth]{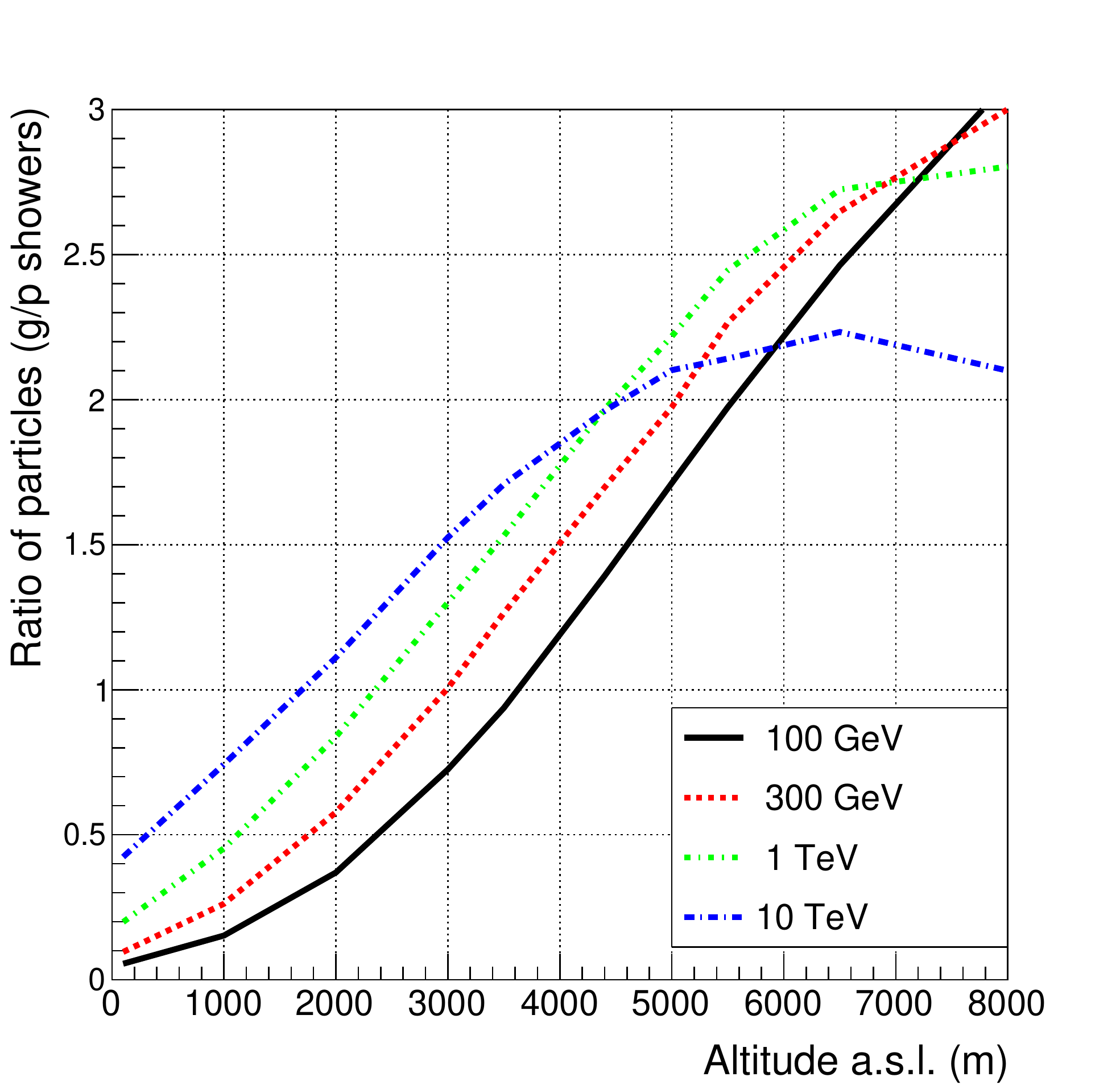} }
\end{minipage}\hfill
\begin{minipage}[t]{.47\linewidth}
  \centerline{\includegraphics[width=\textwidth]{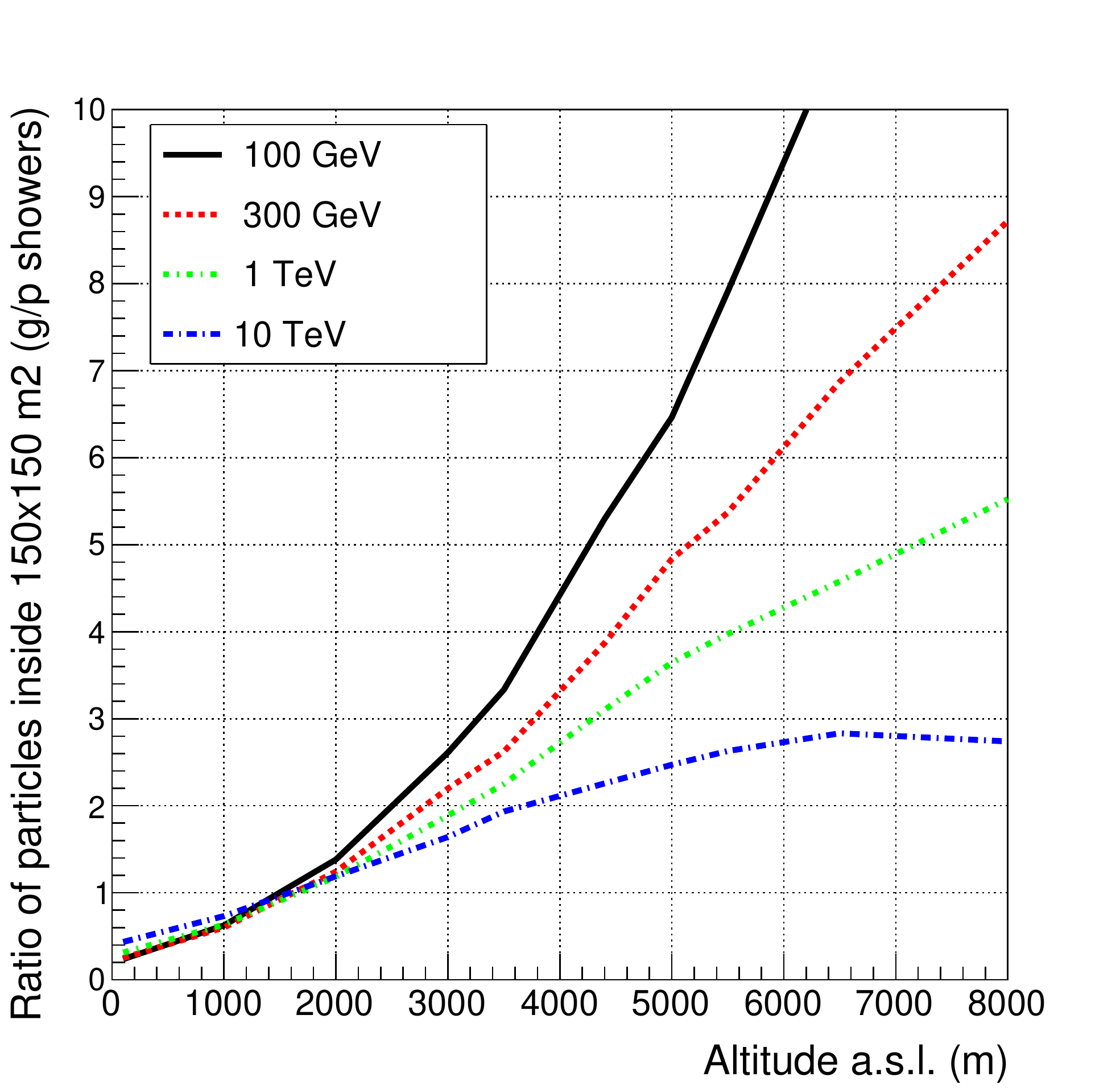} }
\end{minipage}\hfill
\caption[h]{Ratio of particle numbers (charged + photons) in photon- and proton-induced showers of different energies as a function of the observation level. 
The left plot shows the total size, the right one refers to particles contained inside an area 150$\times$150 m$^2$ centered on the shower core \cite{disciascio-icrc2017}.}
\label{fig:ratiosize}
\end{figure}

The small number of charged particles in sub-TeV showers within 150 m from the core imposes to locate experiments at extreme altitudes ($>$4500 m a.s.l.).
At 5500~m a.s.l. 100~GeV $\gamma$-induced showers contain about 8 times more particles than proton showers within 150 m from the core. This fact can be appreciated in the Fig. \ref{fig:ratiosize} where the ratio of particles (charged + photons) in photon- and proton-induced showers of different energies as a function of the observation level are shown.

%
%

\subsection{Relative Trigger Efficiency R}
 
The effective area $A_\text{eff}$ is mainly a function of the number of charged particles at the observation level, the dimension and coverage of the detector and the trigger logic. Moving a given detector at different altitudes $A_\text{eff}$ is proportional to the number of charged particles.

The Fig. \ref{fig:ratiosize} shows the ratio R = N$^{\gamma}$/N$^p$ of secondary particles (charged plus photons) in photon- and proton-induced showers as a function of the altitude. The left plot refers to the total size, the right one to particles contained inside an area 150$\times$150 m$^2$ centered on the shower core.
When R$<$1 the trigger probability is higher for protons than for photons. On the contrary, when R$>$1 the trigger probability, and thus the effective area, of the detector is larger for $\gamma$-showers than for protons, and an intrinsic $\gamma$/hadron-separation is available at higher altitudes. 
Since R is proportional to the ratio of effective areas an altitude $>$4500 m a.s.l. is required to increase the sensitivity of a gamma-ray telescope into the hundreds GeV energy range.

Comparing the two plots of Fig. \ref{fig:ratiosize} we can see that, as expected, the $\gamma$-showers show a more compact particle distribution at the observation level. Therefore, at extreme altitudes the trigger efficiency of photon event at hundreds GeV is highly favoured if we consider only secondary particles within 150 m from the core. 
We note that, as shown in Fig. \ref{fig:size}, the cosmic-ray showers that fake gamma showers are not of the same energies. As an example, at an altitude of 1000 m asl, a 100 GeV proton-induced shower has the same size of a 300 GeV photon shower.

Showers of all energies have the same slope well after the shower maximum: $\approx$1.65x decrease per radiation length (r.l.). This implies that if a given detector is located one radiation length higher in atmosphere, the result will be a $\approx$1.65x decrease of the energy threshold.

But the energy threshold is also a function of the detection medium and of the coverage, the ratio between the detector and instrumented areas. Classical extensive air shower arrays are constituted by a large number of detectors (typically plastic scintillators) spread over an area of order of 10$^4$--10$^5$ m$^2$ with a coverage factor of about 10$^{-3}$. This poor coverage limits the energy threshold because small low-energy showers cannot be efficiently triggered by a sparse array. 
To exploit the potential of the coverage, a high granularity of the read-out must be coupled to image the shower front with high resolution. 

Another important factor to lower the energy threshold of a detector is the secondary photon component detection capability.
Gamma rays dominate the shower particles on ground: at 4300 m asl a 100 GeV photon-induced shower contains on average 7 times more secondary photons than electrons \cite{epas2}. 
In $\gamma$-showers the ratio N$_{\gamma}$/N$_{ch}$ decreases if the comparison is restricted to a small area around the shower core. For instance, we get N$_{\gamma}$/N$_{ch}\sim$3.5 at a distance r $<$ 50 m from the core for 100 GeV showers \cite{epas2}.

\begin{figure}[h!]
\begin{minipage}[t]{.47\linewidth}
  \centerline{\includegraphics[width=\textwidth]{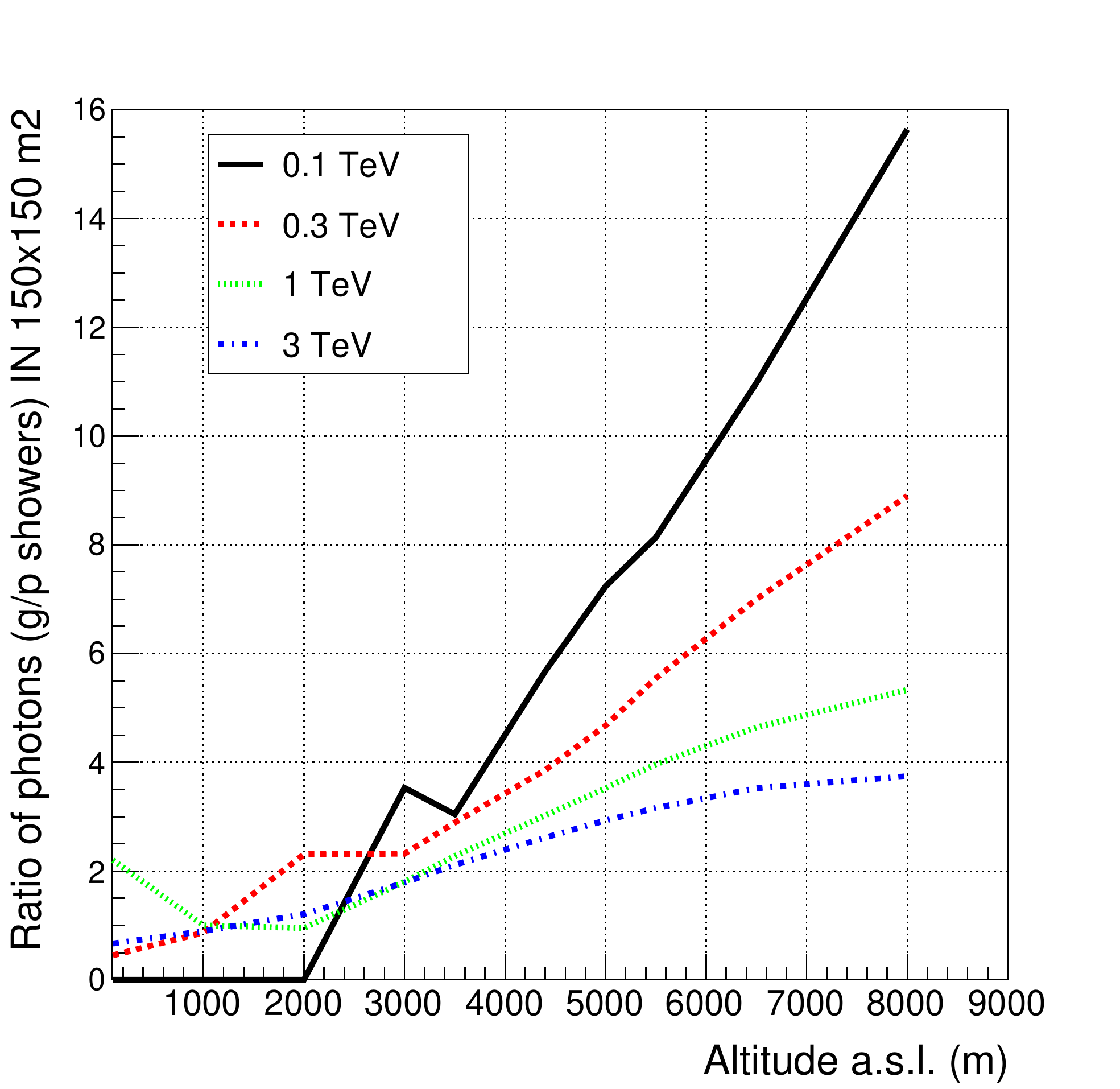} }
  \caption[h]{Ratio of secondary photons in gamma- and proton-induced showers of different energies as a function of the observation level. The particles have been selected inside an area 150$\times$150 m$^2$ centered on the shower core \cite{disciascio-icrc2017}. We note that the kink at about 3000 m asl is only a graphical artifact.} 
\label{fig:photons-pg}
\end{minipage}\hfill
\begin{minipage}[t]{.47\linewidth}
  \centerline{\includegraphics[width=\textwidth]{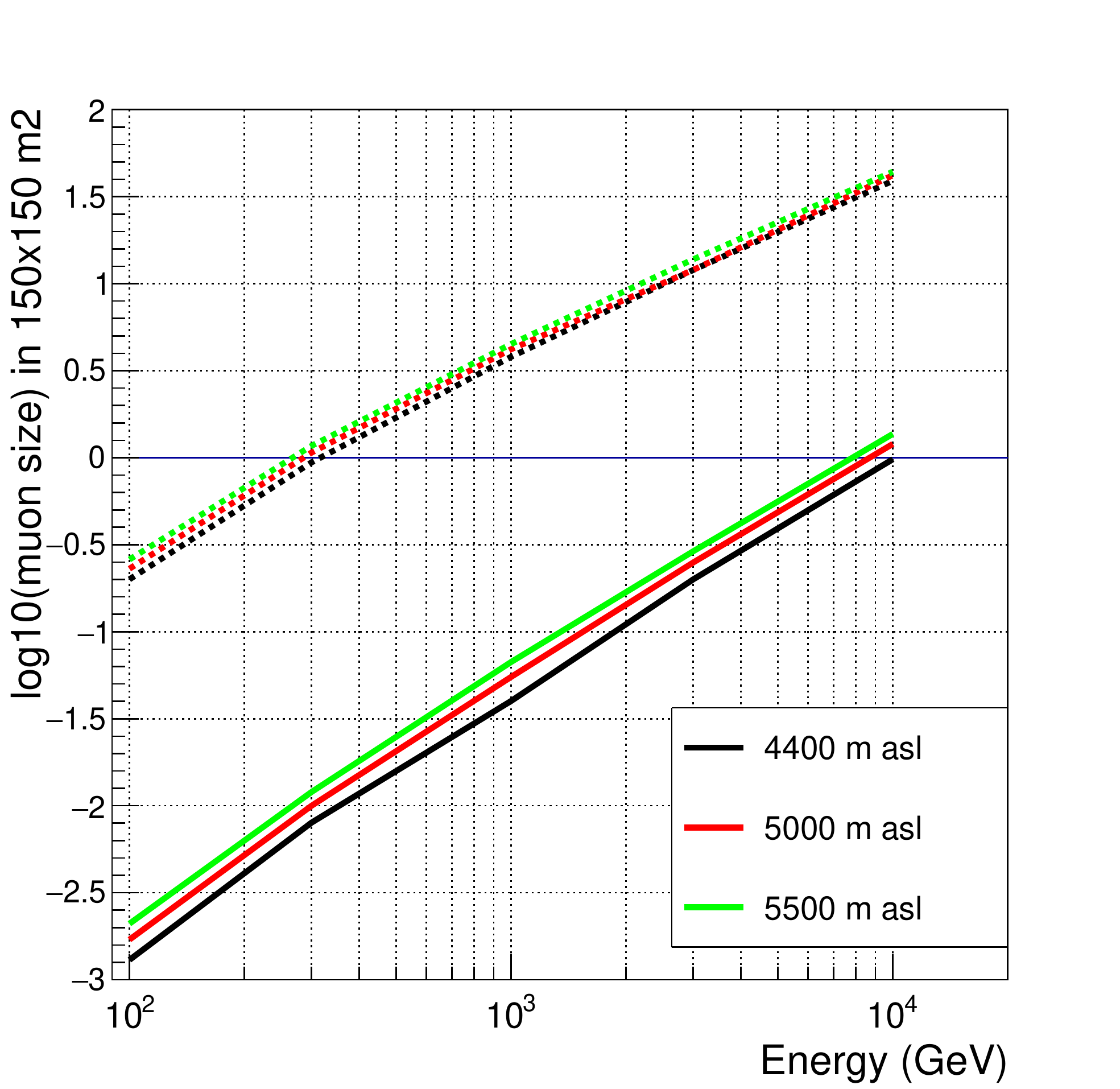} }
\caption[h]{Number of muons within 150 m from the shower core position for proton and photon-induced showers at different altitudes.}
\label{fig:muons}
\end{minipage}\hfill
\end{figure}

In Fig. \ref{fig:photons-pg} the ratio of secondary photons within 150 m from the shower core for gamma- and proton-induced showers of different energies is plotted as a function of the altitude.
The number of secondary photons in low energy $\gamma$-showers exceeds by large factors the number of gammas in p-showers with increasing altitude. 

%
%

Secondary photons can be converted into a electron/positron pair in 2 different ways: 
\begin{enumerate}
    \item[(1)] with 1 radiation length of lead above the counters;
    \item[(2)] with a suitable depth of water in a water Cherenkov detector.
\end{enumerate}
In principle, above 4000 m a.s.l. one radiation length of lead (about 5.5~mm in thickness) allows an increase of the number of charged particles in showers induced by 500~GeV photons up to about 80\%. 

\subsection{The angular resolution}
 
The direction of the primary particle is obtained after reconstructing the time profile of the shower front by using the information from each timing pixel of the experiment. 
Shower particles are concentrated in a front of a nearly spherical shape. A good approximation for particles not far from the shower core is represented by a cone-like shape with an average cone slope of about 0.10 ns/m.
The accuracy in the reconstruction of the shower arrival direction mainly depends on the capability of measuring the relative arrival times of the shower particles. 

The angular resolution of a shower array is a combination of the temporal resolution of the detector unit, the dimension of apparatus, i.e. the dimension of the lever arm in the fitting procedure of the shower front, and the number of temporal hits, i.e. the granularity of the sampling.

\begin{figure}[h!]
\begin{minipage}[t]{.47\linewidth}
  \centerline{\includegraphics[width=\textwidth]{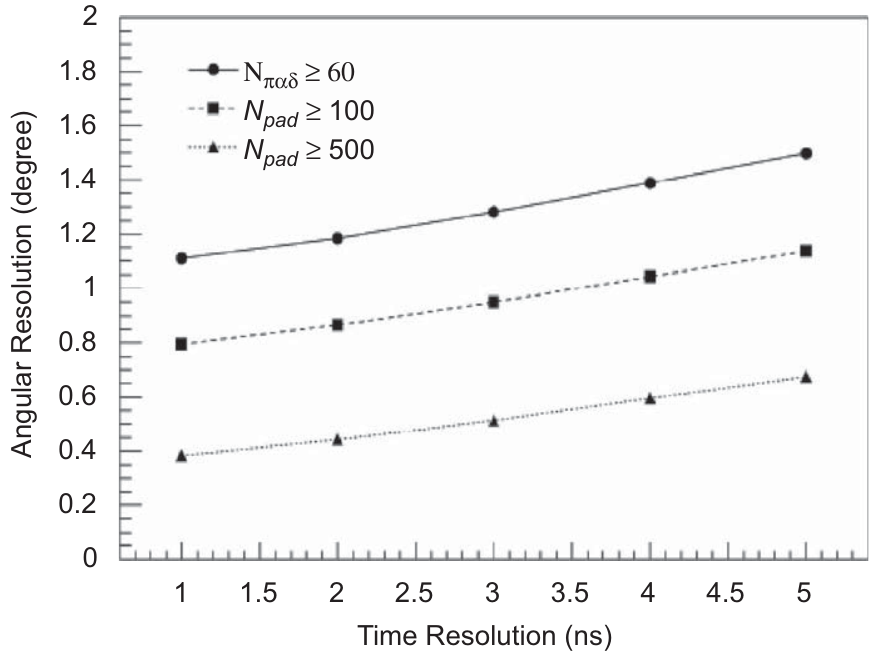} }
\caption[h]{Angular resolution vs time resolutions for RPCs in the ARGO-YBJ experiment. Events with N$_{pad}\geq$60, 100 and 500 fired pads have been selected \cite{aielli2009}. }
\label{fig:angres-timeres}
\end{minipage}\hfill
\begin{minipage}[t]{.47\linewidth}
  \centerline{\includegraphics[width=\textwidth]{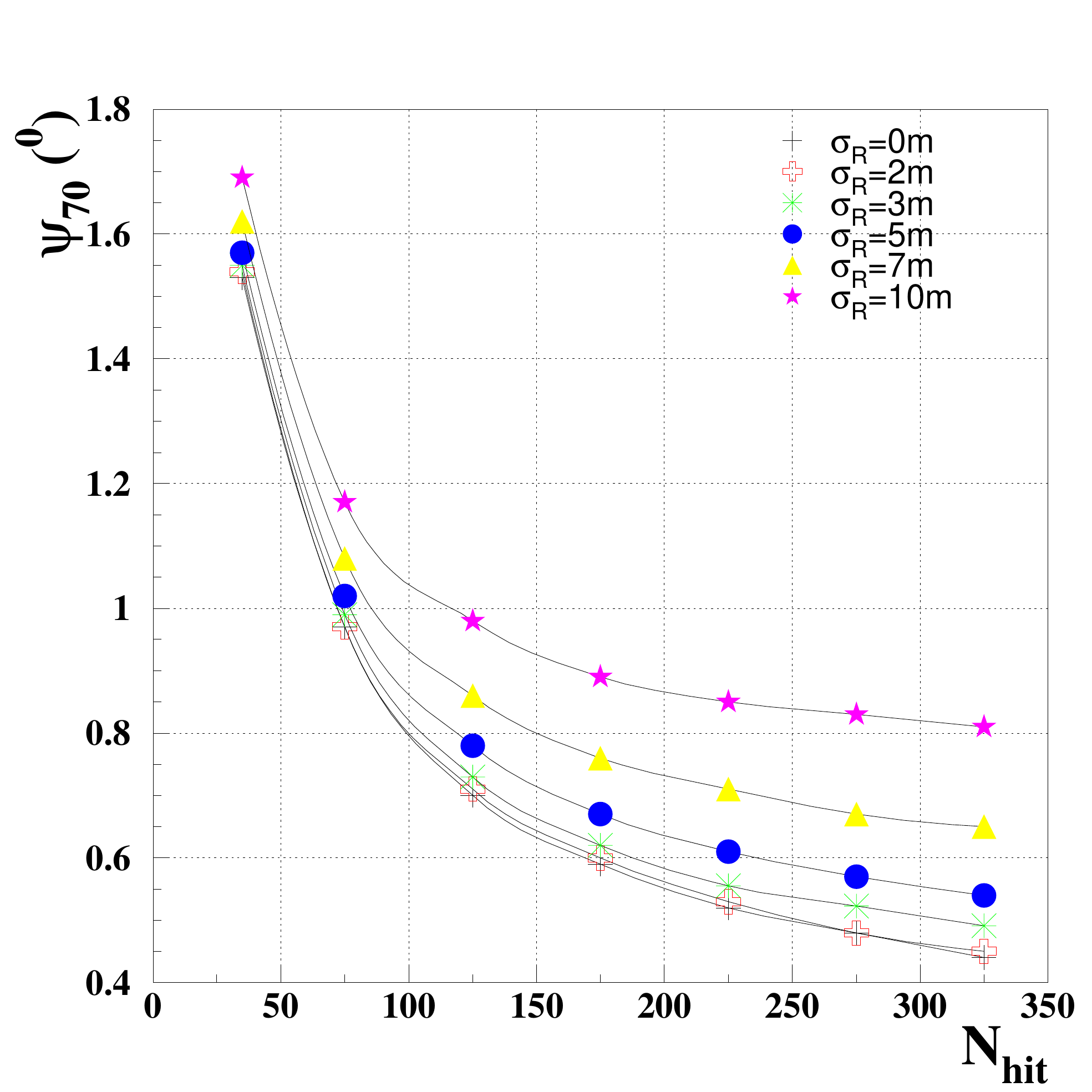} }
\caption[h]{Angular resolution vs shower core position resolution in the ARGO-YBJ experiment.}
\label{fig:angres-corep}
\end{minipage}\hfill
\end{figure}

The time resolution of each detector is determined by the intrinsic time resolution, the propagation time of the signal and the electronic time resolution. As an example, for the ARGO-YBJ experiment the total detector resolution is $\approx$1.3 ns (including RPC intrinsic jitter, strip length, and electronics time resolution).
The dependence of the angular resolution on the time resolution of RPCs in the ARGO-YBJ experiment is shown in Fig. \ref{fig:angres-timeres} \cite{aielli2009}. Events with N$_{pad}\geq$60, 100 and 500 fired pads on the central carpet have been selected.
As it can be seen from the figure, a time resolution in the range between 1 and 2 ns corresponds to a very small change in the angular resolution because the time jitter of the earliest particles in high multiplicity events (>100 hits) is estimated $\approx$1 ns \cite{epas2,argo-test}.

%
%

Following the same arguments as given in \cite{karle1995}, the angular resolution $\sigma_\theta$, averaged on the azimuthal angle $\phi$, is found to depend on multiplicity $N$ and zenith angle $\theta$ as
\begin{equation}
\sigma_\theta\propto\frac{\sigma_t(N)}{\sqrt{N}}\sqrt{\sec \theta}
\end{equation}
where $\sigma_t(N)$ is the average time fluctuation for events with $N$ particles. The factor $\sqrt{\sec \theta}$ accounts for the geometrical effect related to the reduction with increasing $\theta$ of the effective distance between detectors \cite{argo-test}. The dependence of $\sigma_\theta$ upon $N$ is well explained in terms of the combined effect of the time thickness of the extensive air shower disk, as imaged by the detector, and the density of shower particles.

Placing a thin sheet of lead converter (1 radiation length) above the detector (scintillator or RPC) is a well known technique to improve the angular resolution, mainly at the threshold, due to, qualitatively:
\begin{enumerate}
    \item[(1)] absorption of low energy electrons (and photons) which no longer contribute to the time signal;
    \item[(2)] multiplication process of high-energy electrons (and photons) which produce an enhancement of the signal. 
\end{enumerate}
A similar effect is provided by a suitable water depth in a water Cherenkov detector.
The enhanced signal reduces the timing fluctuations: the contributions gained are concentrated near the ideal time profile because the high energy particles travel near the front of the shower while those lost tend to lag far behind.
The ARGO-YBJ experiment has been the only gamma-ray detector operated without a layer of lead above the detectors.
The observation of a number of gamma sources showed the capability of the high granularity sampling provided by the RPC readout in imaging the temporal profile of air showers.

The angular resolution $\sigma_{\theta}$ is related to the opening angle $\Delta \Omega$. If the point spread function of the angular resolution is Gaussian 
\begin{equation}
e^{-\frac{\theta^{2}}{2\sigma_{\theta}^{2}}}
\end{equation}
then the opening angle that maximize the \emph{signal/bkg} ratio is given by $\Delta \theta = 1.58 \sigma_{\theta}$ and it tallies with a fraction of $\epsilon=0.72$ of the events from the direction of the source in the solid angle $\Delta\Omega=2\pi (\cos\Delta\theta)$. 
Therefore, 
\begin{equation}
    \frac{\epsilon(\Delta\Omega)}{\Delta\Omega}\simeq \frac{0.72}{1.6 \sigma_{\theta}}=\frac{0.45}{\sigma_{\theta}} \; \cite{protheroe1984}.
\end{equation} 

The usual method for reconstructing the shower direction is performing a $\chi^2$ fit to the recorded arrival times $t_i$ by minimization of 
\begin{equation}
\chi^2 = \sum_i w ( f - t_i )^2 
\end{equation}
where the sum includes all detectors with a time signal. Usually the function $f$ describes a plane, a cone with a fixed cone slope or a plane with curvature corrections as a function of core distance $r$ and multiplicity $m$. A time offset and two direction cosines are fitted.
The weights $w$ are generally chosen to be an empirical function of the number $m$ of particles registered in a counter, a function of $r$ or a function of $r$ and $m$. This represents in general terms the usual fitting procedure of the "time of flight" technique.
Improvement to this scheme can be achieved by excluding from the analysis the time values belonging to the non-Gaussian tails of the arrival time distributions by performing some successive $\chi^2$ minimizations for each shower \cite{fititer,bartoli2011}.
In fact, the distribution of the arrival times shows non-Gaussian tails at later times, mainly due to multiple scattering of low energy electrons but also to incorrect counter calibrations and to random coincidences. These non-Gaussian tails are expected typically to be  20\% of all measured time values.
 
The resolution in the reconstruction of the shower core position, i.e. the point where the shower axis intersects the detection plane, can affect the angular resolution when functions depending on $r$ are used to describe the temporal profile.
The core position is usually obtained fitting the lateral density distribution of the secondary particles to a modified Nishimura-Kamata-Greisen (NKG) function \cite{llf}.
In Fig. \ref{fig:angres-corep} an example of angular resolution vs shower core position resolution for the ARGO-YBJ carpet is shown. As it can be seen, resolutions worse than about 2 meters affect the angular resolution of the detector.

\begin{figure}[ht]
    \centering
    \vspace{-2.5cm}
    \includegraphics[width=0.9\textwidth]{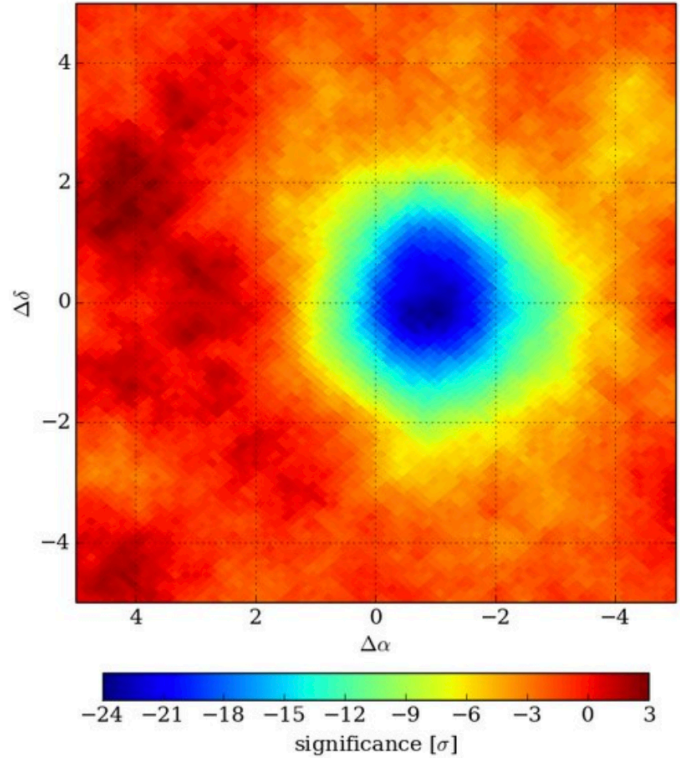}
    \vspace{-3.0cm}
    \caption{Cosmic ray shadow of the Moon in early HAWC data. Note the offset of the shadow to the right, this is due to the Earth's magnetic field deflection of the (charged, primarily proton) cosmic rays. The region to the left provides a limit on the VHE cosmic ray antiproton flux.}
    \label{fig:hawcmoon}
\end{figure}

The standard method to measure the angular resolution and the pointing accuracy of a shower array is to exploit the so-called \emph{Moon shadow} technique which provides unique information on its performance.
CRs blocked in their way to the Earth by the Moon generate a deficit in its direction usually mentioned as ‘‘Moon shadow’’. 
At high energies, the Moon shadow would be observed by an ideal detector as a 0.52$^\circ$ wide circular deficit of events, centred on the Moon position.The actual shape of the deficit as reconstructed by the detector allows the determination of the angular resolution while the position of the deficit allows the evaluation of the absolute pointing accuracy. In addition, charged particles are deflected by the geo-magnetic field by an angle depending on the energy. As a consequence, the observation of the displacement of the Moon shadow at low rigidities can be used to calibrate the relation between the shower size and the primary energy \cite{bartoli2011}. The HAWC moon shadow is shown in Fig. \ref{fig:hawcmoon} and has been used to place constraints on the fraction of antiprotons in the high energy cosmic rays (from the opposite bending from the proton signal seen in the figure) \cite{antiproton}.

\subsection{Background discrimination from the ground}

In 1960 Maze and Zawadzki \cite{maze1960} suggested that in gamma-ray astronomy with shower arrays the background of CRs can be identified and rejected by identifying EAS with an abnormally small number of muons N$_{\mu}$, the so-called \emph{``muon poor''} technique. The existence of such ``unusual'' showers is due to the relatively small photo-nuclear cross section compared with the corresponding value for the proton-nucleus and nucleus-nucleus cross-sections. In gamma showers muons are produced mainly by the photo-production of hadrons 
\begin{equation}
    \gamma + air \to n\pi^{\pm} + m\pi^0 + X (\sigma_{\gamma-air}\sim 1-2 mb),
\end{equation} 
followed by the pion decays in muons and photons, and by muon pair production with a cross section $\sigma_{\gamma-air}\sim$12 $\mu$b. The relevant quantity for the muon content is the total hadronic part of the cross section. In fact, it is the ratio of the total hadronic part of the cross-section to the Bethe-Heitler cross-section one that determines the fraction of the events induced by photons that are hadronic in character. If this ratio is small, then most photon-induced showers must start out an electromagnetic process and no muons are present in the cascade. The probability of pion production, with respect to the probability to produce a $e^+e^-$ pair, is $\sim 3\cdot 10^{-3}$. This implies that the muon content in a gamma shower is only $\sim$10\% the muon content in a shower induced by charged cosmic rays.

For a given shower size N$_e$ and muon number N$_{\mu}$, the selection of muon-poor showers reads
\begin{equation}
    (N_{\mu}/<N_{\mu}>)_{N_e}\leq S_{\mu},
\end{equation}where $< N_{\mu}>$ is the average muon multiplicity expected for a fixed size N$_e$ and $S_{\mu}$ is a optimum threshold value to optimize the sensitivity. In principle $S_{\mu} = <N_{\mu}^{\gamma}>/<N_{\mu}^h>$ where $<N_{\mu}^{\gamma}>$ and $<N_{\mu}^h>$ are the average muon numbers in showers produced by gamma and "normal" (hadron) primaries, respectively. The selection threshold $S_{\mu}$  is not universal but depends on the specific experimental configuration, i.e. the total area of the muon detector and its coverage, the ratio between the sensitive area and the instrumented one. But the key point is the knowledge of the properties of the gamma showers and of the fluctuations of the observed muon number in EAS initiated by photons and primary CRs. In fact, in order to evaluate the rejection power, it is crucial to study how frequently hadronic showers fluctuate in such a way to have a low muon content indistinguishable from gamma-induced events. 

On general ground, there is a close relationship between the adopted model for photo-nuclear interactions and the muon distribution $F(K_{\mu})$ with $K_{\mu} = N_{\mu}/<N_{\mu}>$. As an example, for models assuming a fast increase of the photo-nuclear cross-section, the number of muons expected in high-energy gamma and proton showers become comparable \cite{krys1991}. This imply that the detection of any gamma ray flux with the muon-poor technique would be impossible. 
Therefore, an accurate knowledge of the photo-nuclear process, starting from the cross section, is mandatory. In fact, measurements of the photo-production cross section are limited to $\sqrt{s}\leqslant$200 GeV.

The efficacy of background rejection exploiting the muon content is limited by the number of muons that can be detected. 
According to Monte Carlo simulations, in a proton-induced shower the number of muons is approximately proportional to the
energy of the primary, with about 20 muons above 1 GeV for a 1 TeV proton (200 muons for a 10 TeV muon), but only 4 muons within 150 m of the shower core. As a consequence, the muon poor technique is effective above a few TeV. 
In addition, the fluctuations in the muon number (for a fixed proton energy) are larger than Poisson, with a Gaussian width of $\approx 2.5\sqrt{N_\mu}$, thus there are more events with zero muons than a Poisson calculation. This is an important limiting factor for background discrimination at low energy. Another limiting factor is the high rate of single muons unassociated with any showers at the ground. The need for large full coverage muon detector is evident to exploit the muon poor technique in the TeV energy range.

\begin{figure}[ht]
    \centering
    \vspace{-5.5cm}
    \includegraphics[width=\textwidth]{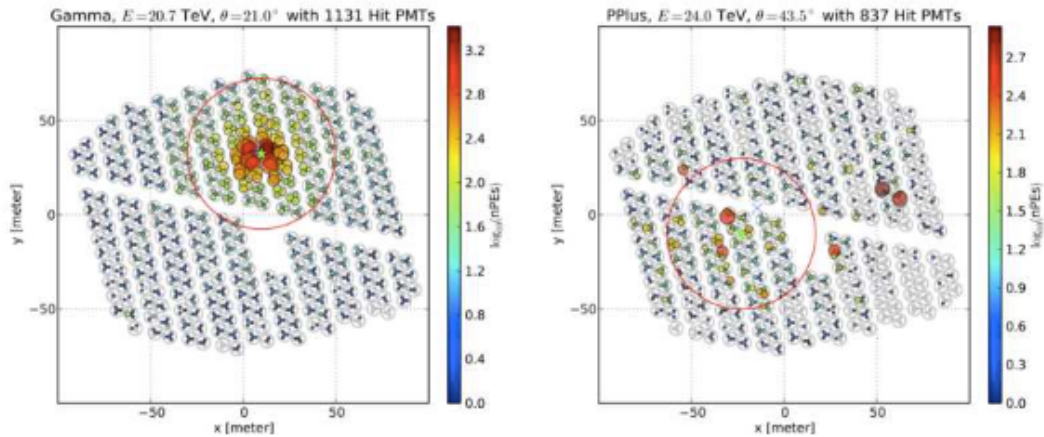}
    \vspace{-6.5cm}
    \caption{HAWC simulation of gamma-ray shower (left panel) and proton shower (right panel) at similar energies. Note the single shower core structure for the electromagnetic cascade for the gamma-ray shower, and the multiple ``centers of gravity'' for the shower core and muon sub-showers in the hadronic simulation.}
    \label{fig:gamma_p}
\end{figure}

The topological differences in muon-poor, or purely electromagnetic showers, and those with muon sub-cascades can be seen, for example, in simulated HAWC gamma and proton results across the array. See \ref{fig:gamma_p} left panel for gamma simulation, note the single core with only random hits across the rest of the array, and the right hand panel for proton simulation where there are multiple strong (muon) hits distinct from the shower core \cite{stefan}. Similar gamma-hadron separation is shown in the HAWC data mentioned earlier, and essential for separation of the photon-induced showers from the cosmic-ray background.

Different experiments in the past (CASA-MIA \cite{borione1997}, EAS-TOP \cite{aglietta1995}, and HEGRA \cite{aharonian2004}) operated muon detectors to measure a possible emission of photons in the 100 TeV energy range, but they set only upper limits mainly due to the small area of the detectors (order of 100 m$^2$) combined with a small efficiency in the muon counting. Only recently, the Tibet AS$\gamma$ \cite{amenomori2019} and HAWC \cite{hawc2020} experiments, detecting muons over areas larger than 1000 m$^2$, reporting evidence of emissions above 100 TeV.
In the last year the breakthrough in VHE gamma-ray astronomy is represented by the LHAASO experiment who observed gamma emission beyond 10$^{15}$ eV opening for the first time the PeV energy range to the observations \cite{cao2021nat}. The large muon detector operated ($\sim$40,000 m$^2$) allows a discrimination at a level of 10$^{-5}$ in the PeV range and a background-free measurement starting from about 100 TeV. 

The background-free regime is very important because in this case the sensitivity is the inverse of the effective area of the array multiplied by the time spent observing a source. Thus, an EAS array with a comparable effective area to a IACT array, with more than one order of magnitude larger time on source, will have a much better sensitivity to the highest energy sources.

\section{Future prospects}
All of the wide field of view experiments mentioned above have been built in the Northern Hemisphere. The construction of a new, wide field of view instrument at sufficiently Southern latitude to continuously monitor the Galactic Center and the inner Galaxy should be a high priority. In order to sensibly find complementarity to the CTA-South IACT effort, this observatory should have the following characteristics:
\begin{enumerate}
    \item[(1)] an energy threshold near 100~GeV, to observe transients;
    \item[(2)] a sensitivity of a few percent of the  Crab level flux below a TeV, for flaring activity detection;
    \item[(3)] an angular resolution around 1$^{\circ}$, to reduce source confusion along the Galactic plane;
    \item[(4)] a real-time trigger facility, and event look-back, to allow multi-messenger astrophysics with CTA, IceCube, LIGO, and other gamma-ray, cosmic-ray, and neutrino experiments;
    \item[(5)] a discrimination against protons at the level of 10$^{-5}$ above 100~TeV to observe the cosmic-ray spectrum knee in the gamma rays; and,
    \item[(6)] an ability to measure the cosmic rays, with some elemental resolution, also up to the knee of the spectrum, to observe the maximum energy of accelerated particles in the cosmic-ray sources and also the anisotropy of the cosmic rays.
\end{enumerate}
Following the ideas above, energy threshold, angular resolution, relative trigger efficiency, effective area, and background rejection, there are a number of efforts in this direction currently under study. These include ALTO, ALPACA, LATTES, STACEX, and SWGO \cite{alto,alpaca,lattes,stacex,swgo}.

ALTO is a follow-on design from HAWC, with an eye towards higher altitudes, denser tank-packing (including hexagonal tanks), faster electronics, and a scintillator panel under the tanks for muon tagging. This effort is a direct push forward on the critical design variables with some innovative ideas on tank construction techniques \cite{alto}.

ALPACA (Andes Large area PArticle detector for Cosmic ray physics and Astronomy) is a project launched in 2016 between Bolivia and Japan aiming at a 83,000~m$^2$ surface air-shower array and a 5,400~m$^2$ underground muon detector array. The site is on a highland pad at an altitude of 4740~m a.s.l. halfway up the road to Mount Chacaltaya on the outskirts of La Paz. The layout of the array is similar to the Tibet AS$\gamma$ experiment, but sited in the Southern Hemisphere \cite{alpaca}.

LATTES (Large Array Telescope for Tracking Energetic Sources) is a proposed hybrid detector utilizing a layer of RPC over a water Cherenkov tank. The base element of the array is a 3~m $\times$ 1.5~m water tank of 0.5~m depth covered by a pair of 1.5~m $\times$ 1.5~m RPCs with a 5.6mm (one radiation length) layer of lead on top. There would be 60 $\times$ 30 of these detector elements comprisong roughly 10,000~m$^2$ of active area installed at 5400~m a.s.l. \cite{lattes}.

STACEX (Southern TeV Astrophysics and Cosmic rays EXperiment) is a similar RPC plus WCDs proposed detector. In this proposed experiment, the RPC panels ``carpet'' the roof of a round, HAWC-like WCD. Total RPC coverage would be 150~m $\times$ 150~m \cite{stacex}.

\begin{figure}[bht]
    \centering
    \vspace{-2cm}
    \includegraphics[angle=270,width=0.9\textwidth]{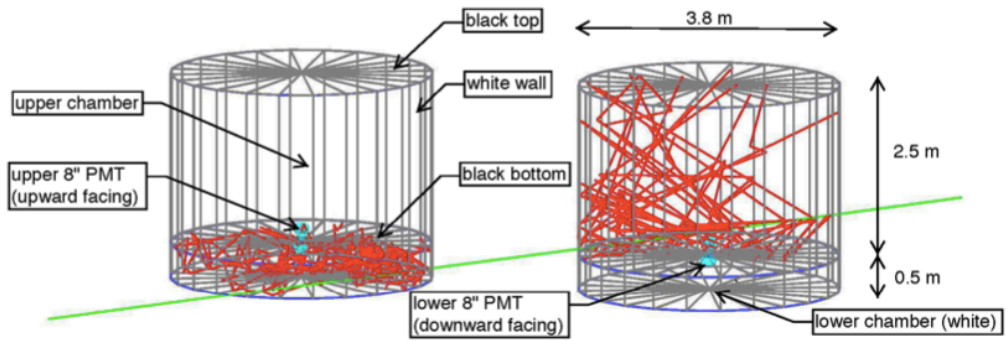}
    \vspace{-2cm}
    \caption{Cylindrical double-layer WCD combing an upper chamber of 3.8~m diameter and 2.5~m depth, with white walls and a black top and bottom, and an entirely white lower chamber of 0.5~m depth. Each chamber has an 8'' PMT, upper facing upwards and lower facing downwards. A simulated muon (green line) is shown passing through both tanks and producing (red track) Cherenkov photons.}
    \label{fig:twolayer}
\end{figure}

SWGO (Southern Wide-field Gamma-ray Observatory) is a consortium (previously called SGSO, Southern Gamma-ray Survey Observatory) examining a very large area water Cherenkov detector employing either a pond or tanks at high altitude in South America \cite{swgo}. Some possible design calls for tanks with two layers for enhanced muon vs. electromagnetic shower discrimination. See Fig. \ref{fig:twolayer} for an illustration of this and more information in the reference \cite{double}.

Outside of the Southern Hemisphere detector concept, there are a number of detector-specific notions which have been raised, often within the context of the aforementioned detectors, which have not yet been implemented in a gamma-ray observatory, but which still merit further investigation:
\begin{itemize}
    \item Neutron detection for cosmic-ray shower rejection; boron-doped plastic scintillator panels are sensitive to up-scattered neutrons from hadronic showers as they hit the ground.
    \item Wavelength-shifting (WLS) fiber readout within a WCD for enhanced photon collection; a ``mop'' of WLS fibers can collect blue Cherenkov photons, convert them to green light, and waveguide that light to smaller photo detectors.
    \item Fast FPGA-based photon correlator allowing for low signal level, few sensor hit, reconstruction; for example, search for all 5-10 PMT hits which point to the Crab, along with a location at the same zenith but differing orientation for background subtraction. Sometimes called a ``vector telescope.''
    \item Large area photodetectors potentially built up from Silicon PMs; lower cost, lower voltage, and lower transit time spread are possible.
    \item Scintillator additions within the WCD; liquid scintillator or quantum dots to enhance photon yield, and potentially help with particle ID as well.
    \item Freezing point depression of the fluid in a WCD; to allow for no freezing even at the highest South American site locations.
    \item Low power electronics; improving the sustainable footprint of the detector, potentially combined with the use of renewable energy, recyclable detector elements, and a minimally invasive installation at the site.
\end{itemize}
New detector designs typically are fairly conservative, improving incrementally on previous designs, especially in a fairly mature design environment such as the ground-based VHE gamma-ray observatory. The proposed Southern Hemisphere telescopes reflect that path but do allow for the exploration of other potential improvements. LHAASO represents the current state-of-the-art pushing the design primarily in scale and the aggressive combination of multiple detector techniques rather than novel technologies. 


 

\expandafter\ifx\csname url\endcsname\relax
  \def\url#1{{\tt #1}}\fi
\expandafter\ifx\csname urlprefix\endcsname\relax\def\urlprefix{URL }\fi
\providecommand{\eprint}[2][]{\url{#2}}

\end{document}